\documentclass[prl,twocolumn,aps,showpacs]{revtex4-1}

\usepackage{epsfig}
\usepackage{graphicx}
\usepackage{color}

\begin{document}

\title{Temperature dependence of a vortex in a superfluid Fermi gas}

\author{S. Simonucci, P. Pieri, and G. C. Strinati}

\affiliation{Division of Physics, School of Science and Technology \\ Universit\`{a} di Camerino, 62032 Camerino (MC), Italy \\
                 and \\ INFN, Sezione di Perugia, 06123 Perugia (PG), Italy}

\date{\today}

\begin{abstract}
The temperature dependence of an isolated quantum vortex, embedded in an otherwise homogeneous fermionic superfluid of infinite extent, is determined via the Bogoliubov-de Gennes (BdG) equations across the BCS-BEC crossover. 
Emphasis is given to the BCS side of this crossover, where it is physically relevant to extend this study up to the critical temperature for the loss of the superfluid phase, such that the size of the vortex increases without bound. 
To this end, two novel techniques are introduced. 
The first one solves the BdG equations with ``free boundary conditions'', which allows one to determine with high accuracy how the vortex profile matches its asymptotic value at a large distance from the center, thus avoiding a common practice of constraining the vortex in a cylinder with infinite walls. 
The second one improves on the regularization procedure of the self-consistent gap equation when the inter-particle interaction is of the contact type, and permits to considerably reduce the time needed for its numerical integration, by drawing elements from the derivation of the Gross-Pitaevskii equation for composite bosons starting from the BdG equations. 
\end{abstract}

\maketitle

\section{I. Introduction} 
\label{sec:introduction}

Vortices are at the essence of superfluidity and of its deep connection with Bose-Einstein condensation (BEC) \cite{PiSt-2003}.
They have thus received considerable interest with the raise of ultra-cold dilute trapped Bose gases \cite{DGPS-1999}, where they can be generated by 
setting the trap into rotation \cite{Fetter-2009} and have been the subject of experimental investigation \cite{Ketterle-2001}.
In this context, isolated vortices or even vortex arrays have mainly been studied theoretically in terms of the Gross-Pitaevskii (GP) equation 
for the wave function of the condensate \cite{Gross-1961,Pitaevskii-1961}, which was specifically introduced to describe an isolated vortex in an otherwise uniform dilute Bose-Einstein condensate.

Subsequent interest in ultra-cold dilute trapped Fermi gases and in the associated BCS-BEC crossover \cite{BDZ-2008,GPS-2008} (whereby a continuos evolution is achieved from a BCS-like situation with highly overlapping Cooper pairs, to a BEC-like situation where composite bosons form out of fermion pairs and condense at sufficiently low temperature) has raised the issue of the description of vortices in Fermi systems, for which the Pauli principle requires one to consider in general a whole set of one-particle wave functions instead of a single condensate wave function.
In this context, isolated vortices (or even vortex arrays) have been studied theoretically in terms of the Bogoliubov-de Gennes (BdG) equations
\cite{BdG}, which were introduced as an extension of the BCS approach \cite{Schrieffer-1964} to describe a non-uniform Fermi superfluid. 
Experimentally, arrays of vortices have been detected throughout the BCS-BEC crossover once trapped Fermi atoms were set into rotation \cite{Ketterle-2005}.

From the computational side, solution of the BdG equations for the fermionic wave functions is much more involved and time consuming than the solution of the GP equation for the bosonic condensate wave function.
For this reason, consideration has essentially been limited to the study of an isolated vortex (with the exception of arrays of vortices in the weak-coupling (BCS) limit \cite{Feder-2005,Castin-2006}).
In particular, an isolated vortex was considered by solving the BdG equations in Refs.\cite{SRH-2006} and \cite{Levin-2006} at zero temperature throughout the BCS-BEC crossover, and in Ref.\cite{Bruun-2004} at finite temperature but in the weak-coupling (BCS) limit only.
In these works, the superfluid was enclosed in a cylinder of radius $R$.

Aim of the present paper is to extend the calculation of the fermionic BdG equations for a single vortex over the \emph{whole} temperature range from zero 
up to the critical temperature $T_{c}$ for the loss of the superfluid phase, while spanning at the same time the entire BCS-BEC crossover.
In practice, the crossover between the BCS and BEC regimes is essentially exhausted within a range $\approx 1$ about the unitary limit at $(k_{F}a_{F})^{-1} = 0$ where the scattering length $a_{F}$ of the two-fermion problem diverges ($k_{F}$ being the Fermi wave vector related to the bulk density $n_{0}$ via
$n_{0} = k_{F}^{3}/3 \pi^{2}$).

\begin{figure}[t]
\includegraphics[angle=0,width=7.5cm]{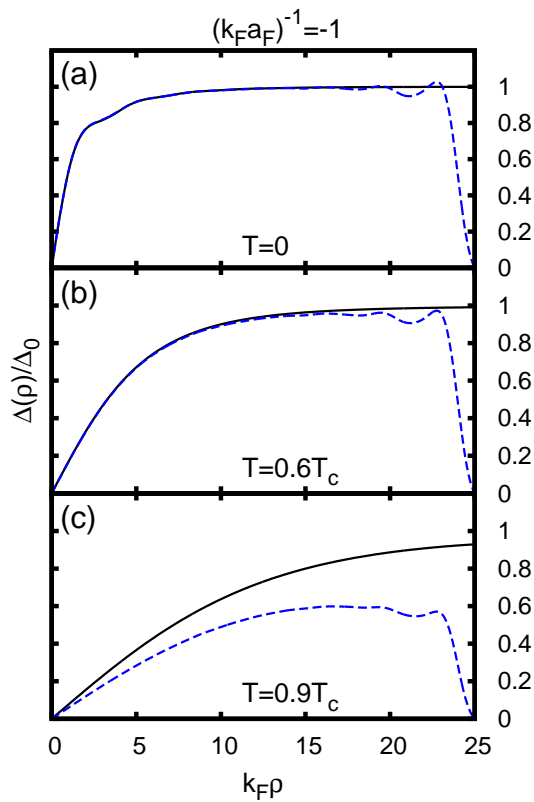}
\caption{Profile of the order parameter $\Delta(\rho)$ (normalized to its asymptotic value $\Delta_{0}$ at the given temperature) for an isolated vortex versus the distance $\rho$ from the center. The weak-coupling case with $(k_{F}a_{F})^{-1} = -1$ is considered for three different temperatures: (a) $T=0$; (b) $T=0.6 T_{c}$; (c) $T=0.9 T_{c}$. In the three cases, the calculation using ``free boundary conditions'' (full line) is compared against that using a cylindrical box (dashed line), with the value $R = 25 k_{F}^{-1}$ for the radius as typically taken in previous calculations \cite{SRH-2006}.}
\label{Figure-1}
\end{figure}

This will require us to avoid constraining the superfluid within a cylinder of radius $R$ with rigid walls, but to let it be free of expanding its size without bound when approaching $T_{c}$ from below.
To this end, appropriate ``free boundary conditions'' will have to be implemented for the BdG equations, in order to recover their correct asymptotic solution far away from the center of the vortex when its size would exceed any reasonable value one could take for $R$.
The advantage of avoiding the use of a finite value $R$ can be perceived, in practice, even somewhat away from $T_{c}$, as it can be seen from the 
weak-coupling case reported in Fig.÷\ref{Figure-1} for the sake of example.

In this way, we will be able to obtain the \emph{healing length} for an isolated vortex as a function of the temperature $T$ from $T=0$ up to (quite close 
to) $T_{c}$ \emph{and} of the coupling parameter $(k_{F}a_{F})^{-1}$ spanning the BCS-BEC crossover.

This information about the way the superfluid healing length can be fine-tuned in a Fermi gas, by varying not only the temperature but also the inter-particle coupling (or both), may also be relevant for the emerging field of superfluid interferometers \cite{PT-2012}, in the case it could be possible to realize them in practice by coupling systems of ultra-cold dilute trapped Fermi atoms.

In the course of the present calculation, we shall also improve on the regularization procedure of the self-consistent gap equation which was used in the literature for similar problems \cite{BY-2002,GU-2003,LHD-2007} and is required when, like in the present context, the inter-particle interaction is of the contact type.
This will permits us to reduce considerably the computational time needed for the numerical integration of the BdG equations, while leaving unaltered the numerical accuracy.
To this end, elements will be drawn from the derivation of the GP equation for composite bosons that was obtained in Ref.\cite{PS-2003} on the BEC side of the crossover starting from the BdG equations.

A second, yet not less important, purpose of the present paper is to obtain \emph{as accurate as possible} numerical solutions of the BdG equations for a non-trivial physical problem (like that of an isolated vortex embedded in an infinite superfluid) under a wide variety of circumstances.
These numerical solutions could, in fact, be used in the future as a ``benchmark'' for the results obtained alternatively by solving approximate local (differential) equations for the gap parameter, which could take the place of the fermionic BdG equations at least in some approximate sense.
In turn, these local equations should be better suited to deal with more complex problems like the arrays of vortices and the moment of inertia of the superfluid, which can be explored experimentally with ultra-cold trapped Fermi atoms \cite{Ketterle-2005,Grimm-2011} but remain too difficult to be approached theoretically by solving directly the fermionic BdG equations.

As an example, the GP equation for composite bosons at low temperature that was derived in Ref.\cite{PS-2003} has already been tested in the context of the Josephson effect \cite{SPS-2010}, to produce results quite similar to those obtained by solving the BdG equations on the BEC side of the crossover, albeit in a much more efficient way.
Similarly, the Ginzburg-Landau (GL) equation for Cooper pairs, that was derived by Gorkov \cite{Gorkov-1959} on the BCS side of the crossover \emph{and} close to $T_{c}$ also starting from the fermionic BdG equations, can be most readily applied to non-uniform superconductors under a variety of circumstances \cite{Tinkham-1975} since its solution is considerably simpler than that of the original BdG equations.

Along these lines, attempts have already been made in the past to derive from the BdG equations extensions of the GL equation, which would still apply to the weak-coupling (BCS) regime but somewhat deeper in the superfluid phase away from the vicinity to $T_{c}$ 
\cite{Tewordt-1963,Werthamer-1963,Werthamer-1964}.
More recently, a systematic expansion of the BdG equations in terms of the small parameter $(T_{c} -T)/T_{c}$ was considered in the weak-coupling regime,
but was explicitly tested for the spatially uniform case only \cite{Shanenko-2011}.
A satisfactory test of the above (as well of other) proposal for differential equations, that aim at extending the validity GL equation deep in the superfluid region, is thus apparently still pending and the accurate solution of the BdG equations we obtain in the present paper may provide the awaited ground for this comparison.

In this context, an additional important information that can be obtained by the present approach comes from the analysis of how alternative energy ranges in the solution of the BdG equations (namely, bound states, and near and far continuum) contribute to the different spatial regions in which the profiles of physical quantities associated with a vortex (like the gap parameter itself and the number and current densities) can be partitioned.
This kind of information is, in fact, expected to be relevant in future work in order to assess the validity of approximate local (differential) equations for the gap parameter.
 
The paper is organized as follows.
Section II considers the solution of the BdG equations for an isolated vortex embedded in an infinite superfluid, for which ``free boundary conditions'' are introduced and the associated normalization of the wave functions in the continuum is obtained.
The spatial profiles of the vortex obtained in this way under a variety of circumstances are reported in Section III.
Section IV discusses the procedure through which the healing length of the vortex, as a function of coupling and temperature, can be extracted from the above profiles.
Section V provides an analysis of the contribution of the different energy ranges in the BdG equations to different portions in the profiles of physical 
quantities.
Section VI gives our conclusions.
The way the boundary conditions are implemented is discussed in detail in Appendix A, the improved regularization procedure for the gap equation is derived in Appendix B, and the related expressions for the number and current densities are reported in Appendix C. 

\section{II. Solution of the Bogoliubov-de Gennes equations with free boundary conditions} 
\label{sec:BdG_equations}

In this Section, we discuss in detail the solution of the fermionic BdG equations in cylindrical coordinates for an isolated vortex embedded in an otherwise infinite superfluid.
To be able to deal with situations when the size of the vortex grows without bound upon approaching $T_{c}$ (in practice, when it exceeds a few dozens times $k_{F}^{-1}$), an explicit numerical integration of the BdG equations will be performed from the center of the vortex outwards only in a limited radial range, at the boundary of which connection with asymptotic solutions will be sought in terms of known functions of mathematical physics.
Knowledge of these asymptotic solutions will also enable us to to determine the normalization of the eigen-solutions of the continuum part of the spectrum of the BdG equations.
This step is of particular importance, since it turns out that the continuum part of the spectrum exhausts in practice most part of the contribution to the relevant physical quantities.

\vspace{0.05cm}
\begin{center}
{\bf A. BdG equations for an isolated vortex embedded in an infinite medium}
\end{center}
\vspace{0.05cm}

The fermionic BdG equations read:

\begin{equation}
\left( 
\begin{array}{cc}
\mathcal{H}(\mathbf{r}) & \Delta(\mathbf{r})            \\
\Delta(\mathbf{r})^{*}  & - \mathcal{H}(\mathbf{r})  
\end{array} 
\right)
\left( \begin{array}{c}
u_{\nu}(\mathbf{r}) \\
v_{\nu}(\mathbf{r}) 
\end{array} 
\right) 
= \varepsilon_{\nu}
\left( \begin{array}{c}
u_{\nu}(\mathbf{r}) \\
v_{\nu}(\mathbf{r}) 
\end{array} 
\right)                                          \label{BdG-equations} 
\end{equation}
\noindent
where $\mathcal{H}(\mathbf{r}) = - \nabla^{2}/2m - \mu$ ($m$ being the fermion mass, $\mu$ the chemical potential, and $\hbar = 1$ troughout).
The local gap parameter $\Delta(\mathbf{r})$ is determined via the \emph{self-consistent condition}:
\begin{equation}
\Delta(\mathbf{r}) = - v_{0} \sum_{\nu} u_{\nu}(\mathbf{r}) v_{\nu}(\mathbf{r})^{*} \left[ 1 - 2 f_{F}(\varepsilon_{\nu}) \right]       
\label{self-consistency}
\end{equation}
\noindent
where $f_{F}(\epsilon)=(e^{\epsilon/(k_{B}T)} +1)^{-1}$ is the Fermi function at temperature $T$ ($k_{B}$ being Boltzmann constant) and 
$v_{0}$ is the (bare) coupling constant of the contact interaction. Only positive values of the eigenvalues $\epsilon_{\nu}$ can be explicitly
considered \cite{BdG}.

We are specifically interested in a spatially dependent gap parameter $\Delta(\mathbf{r})$ with cylindrical symmetry
\begin{equation}
\Delta(\mathbf{r}) = \Delta(\rho,\varphi,z) = \Delta(\rho) \, e^{i n \varphi}                                          \label{Delta-cylindrical-symmetry}
\end{equation}
\noindent
that corresponds to an isolated vortex directed along the $z$ axis with circulation quantum $n$ ($n$ integer). [We shall take $n = 1$ eventually.]
The associated wave functions of Eqs.(\ref{BdG-equations}) have the form:
\begin{eqnarray}
u_{\nu,\ell,k_{z}}(\mathbf{r}) & = & u_{\nu}(\rho) \, e^{i \ell \varphi} \, e^{i k_{z} z}                             \nonumber \\
v_{\nu,\ell,k_{z}}(\mathbf{r}) & = & v_{\nu}(\rho) \, e^{i (\ell - n) \varphi} \, e^{i k_{z} z}                      \label{cylindrical-wave-functions}
\end{eqnarray}
\noindent
($\ell$ integer) where $\Delta(\rho)$, $u_{\nu}(\rho)$, and $v_{\nu}(\rho)$ are real  functions.
The BdG equations (\ref{BdG-equations}) then become:
\begin{eqnarray}
\mathcal{O}_{\ell} \, u_{\nu}(\rho) + \Delta(\rho) v_{\nu}(\rho) & = & \varepsilon_{\nu} u_{\nu}(\rho)               \nonumber \\
- \mathcal{O}_{\ell -n} \, v_{\nu}(\rho) + \Delta(\rho) u_{\nu}(\rho) & = & \varepsilon_{\nu} v_{\nu}(\rho)         
\label{cylindrical-BdG-equations}
\end{eqnarray}
\noindent
involving the radial operator 
\begin{equation}
\mathcal{O}_{\ell} = \left[ - \frac{1}{2 m \rho} \frac{d}{d \rho} \left( \rho \frac{d}{d \rho} \right) + \frac{\ell^{2}}{2 m \rho^{2}} - \tilde{\mu} \right] \label{O_ell-definition}
\end{equation}
\noindent
where $\tilde{\mu} = \mu - k_{z}^{2}/2 m$ is the reduced chemical potential.

Each of the two second-order differential equations (\ref{cylindrical-BdG-equations}) admits a regular solution in $\rho = 0$, which behave respectively as
$u_{\nu}(\rho) \sim \rho^{|\ell|}$ and $v_{\nu}(\rho) \sim \rho^{|\ell - n|}$.
In particular, for $n = 1$ (whereby $\Delta(\rho) = \eta \rho$ for $\rho \rightarrow 0$ with $\eta$ constant), two \emph{independent} solutions of the coupled equations (\ref{cylindrical-BdG-equations}) can be obtained by taking the \emph{indicial conditions}:
\begin{eqnarray}
u_{\nu}^{(1)}(\rho) & = & \rho^{|\ell|} \, + \, \cdots                            \nonumber \\
v_{\nu}^{(1)}(\rho) & = & \beta \rho^{|\ell|+3} \, + \, \cdots                \label{indicial-conditions-1}
\end{eqnarray}
\noindent
where $(4 + 3|\ell| + \ell) \beta /m + \eta = 0$, and
\begin{eqnarray}
u_{\nu}^{(2)}(\rho) & = & \gamma \rho^{|\ell-1|+3} \, + \, \cdots            \nonumber \\
v_{\nu}^{(2)}(\rho) & = & \rho^{|\ell-1|} \, + \, \cdots                        \label{indicial-conditions-2}
\end{eqnarray}
\noindent
where $[\ell^{2} - (|\ell - 1| + 3)^{2}] \gamma /(2m) + \eta = 0$.

The differential equations (\ref{cylindrical-BdG-equations}) are integrated numerically from $\rho = 0$ up to an outer value $R_{\mathrm{out}}$, and for several values of $\ell$ up to a maximum value $\ell_{\mathrm{max}}$.
Here, $\ell_{\mathrm{max}}$ and $R_{\mathrm{out}}$ can be related to each other as follows:

\noindent
(i) To begin with, one selects a cutoff energy $E_{c}$ such that only (positive) eigenvalues $\varepsilon_{\nu}$ up to $E_{c} - \mu$ are explicitly considered in the solution of Eqs.(\ref{cylindrical-BdG-equations}) (the remaining eigenvalues larger than $E_{c} - \mu$ will be dealt with separately by the regularization procedure for the gap equation described in Appendix B);

\noindent
(ii) One then chooses a value of $R_{\mathrm{out}}$ such that for $\rho > R_{\mathrm{out}}$ the gap $\Delta(\rho)$ in Eqs.(\ref{cylindrical-BdG-equations}) has reached its asymptotic (bulk) value $\Delta_{0}$, say, within $1 \%$ (values $k_{F} R_{\mathrm{out}} \simeq 60 \div 200$ prove sufficient for all practical purposes);

\noindent
(iii) Finally, one solves numerically Eqs.(\ref{cylindrical-BdG-equations}) for values of $\ell$ up to $\ell_{\mathrm{max}}$ such that 
$\ell_{\mathrm{max}}^{2}/(2mR_{\mathrm{out}}^{2}) \sim E_{c}$, that is to say, $\ell_{\mathrm{max}} \sim k_{c}R_{\mathrm{out}}$ with $k_{c}=\sqrt{2mE_{c}}$
(in practice, we have taken $\ell_{\mathrm{max}}$ not smaller than $200$).

It is clear that a reasonable estimate of the value of $R_{\mathrm{out}}$ entails knowledge of the profile of $\Delta(\rho)$, which in turn requires the solution of the self-consistent condition (\ref{self-consistency}).
We defer to Appendix B the solution of Eq. (\ref{self-consistency}) together with a proper treatment of the convergence of the sum over $\nu$ for large values 
of $\varepsilon_{\nu}$.
In this context, a novel regularization procedure for the gap equation (\ref{self-consistency}) will be introduced, which improves on regularization procedures previously considered in the literature \cite{BY-2002,LHD-2007} (thereby effectively reducing the numerical value of $E_{c}$).

\vspace{0.05cm}
\begin{center}
{\bf B. Asymptotic behavior of the wave functions}
\end{center}
\vspace{0.05cm}

For an isolated vortex embedded in an otherwise infinite superfluid medium, the eigenvalues $\varepsilon_{\nu}$ of the BdG equations (\ref{cylindrical-BdG-equations})
belong to a \emph{continuous spectrum} above the threshold $\Delta_{0}$ (apart from the Andreev-Saint-James bound states that lie below this threshold).
For the wave functions belonging to this continuum, in turn, the normalization is determined from their ``asymptotic'' behavior for large values of $\rho$,
which may be identified only for $\rho \gg R_{\mathrm{out}}$.
For this reason, the asymptotic behavior of $u_{\nu}(\rho)$ and $v_{\nu}(\rho)$ for $\rho \rightarrow \infty$ has eventually to be searched in terms of known functions of mathematical physics, which is however \emph{not} possible for the radial BdG equations (\ref{cylindrical-BdG-equations}) as they stand.

To overcome this problem, we have adopted the following strategy.
If $R_{\mathrm{out}}$ is large enough, the centrifugal terms $\ell^{2}/(2 m \rho^{2})$ and $(\ell-n)^{2}/(2 m \rho^{2})$ in Eqs.(\ref{cylindrical-BdG-equations})
are important only for large values of $\ell$, in such a way that we may replace $\ell$ and $(\ell-n)$ by their \emph{average value}:
\begin{equation}
\ell' = \frac{\left[ \ell + (\ell - n) \right]}{2} = \ell - \frac{n}{2} \, \longrightarrow \, \ell - \frac{1}{2} \, .      \label{l'}
\end{equation}
\noindent
By this replacement, in the centrifugal terms for $\rho > R_{\mathrm{out}}$ we make an error smaller than $E_{c}/\ell_{\mathrm{max}}$.
Accordingly, for $\rho \ge R_{\mathrm{out}}$ in the place of Eqs.(\ref{cylindrical-BdG-equations}) we consider the following ``modified'' BdG equations with 
\emph{a common value} of $\ell'$:
\begin{eqnarray}
\mathcal{O}_{\ell'} \, u_{\nu}(\rho) + \Delta_{0} \, v_{\nu}(\rho)   & = & \varepsilon_{\nu} \, u_{\nu}(\rho)              \nonumber \\
- \mathcal{O}_{\ell'} \, v_{\nu}(\rho) + \Delta_{0} \, u_{\nu}(\rho) & = & \varepsilon_{\nu} \, v_{\nu}(\rho) \, .
\label{modified-cylindrical-BdG-equations}
\end{eqnarray}
\noindent
These coupled equations can be solved analytically in terms of known functions of of mathematical physics, by considering the auxiliary equation:
\begin{equation}
\left[ - \frac{1}{2 m \rho} \frac{d}{d \rho} \left( \rho \frac{d}{d \rho} \right) + \frac{\ell'^{2}}{2 m \rho^{2}} - \tilde{\mu} \right] f_{k}(\rho) =
\left(\frac{k^{2}}{2m} - \mu \right) f_{k}(\rho)                                                 \label{auxiliary-equation}
\end{equation}
\noindent
where $k^{2} = k_{\perp}^{2} + k_{z}^{2}$.
This equation is equivalent to the canonical equation of the Bessel functions of index $|\ell'|$ 
\begin{equation}
\zeta^{2} \, \frac{d f(\zeta)}{d\zeta^{2}} + \zeta \, \frac{d f(\zeta)}{d\zeta} + (\zeta^{2} - \ell'^{2}) f(\zeta) = 0   
\label{equation-Bessel-functions}
\end{equation}
\noindent
in the dimensionless variable $\zeta=k_{\perp} \rho$ \cite{AS-1972}.
The solutions to Eqs.(\ref{modified-cylindrical-BdG-equations}) are thus sought in the form
\begin{equation}
u_{\nu}(\rho) = u_{k} \, f(k_{\perp}\rho) \hspace{0.3cm} , \hspace{0.3cm} v_{\nu}(\rho) = v_{k} \, f(k_{\perp}\rho)  \hspace{0.3cm} ,   
\label{smart-solution}
\end{equation}
\noindent
which reduce Eqs.(\ref{modified-cylindrical-BdG-equations}) to the standard system of 
algebraic equations \cite{BdG}
\begin{eqnarray}
\left(\frac{k^{2}}{2m} - \mu \right) \, u_{k} + \Delta_{0} \, v_{k}   & = & \varepsilon_{k} \, u_{k}              \nonumber \\
- \left(\frac{k^{2}}{2m} - \mu \right) \, v_{k} + \Delta_{0} \, u_{k} & = & \varepsilon_{k} \, v_{k} \, ,         \label{algebraic-BdG-equations}
\end{eqnarray}
\vspace{-0.5cm}
\noindent
yielding
\begin{equation}
u_{k}^{2} = \frac{1}{2} \left( 1 \, + \, \frac{\frac{k^{2}}{2m} - \mu}{\varepsilon_{k}} \right) = 1 - v_{k}^{2} \label{u-v-solutions}
\end{equation}
\vspace{-0.2cm}
\noindent
where 
\begin{equation}
\varepsilon_{k} = \sqrt{ \left( \frac{k^{2}}{2m} - \mu \right)^{2} + \Delta_{0}^{2} } =
\sqrt{ \left( \frac{k_{\perp}^{2}}{2m} - \tilde{\mu} \right)^{2} + \Delta_{0}^{2} } \,\, .           \label{energy-eigenvalue}
\end{equation}

\vspace{0.4cm}
When dealing with the continuum spectrum, it is convenient to use the energy eigenvalue $\varepsilon$ as the independent variable.
This constrains $k_{\perp}$ in Eq.(\ref{energy-eigenvalue}) to the values:
\begin{equation}
k_{\perp} \, = \, \pm \, \sqrt{2m \tilde{\mu} \, \pm \, 2m \, \sqrt{ \varepsilon^{2} - \Delta_{0}^{2}} }     \label{k-perp_vs_energy}
\end{equation}
\noindent
for given $\varepsilon$ and $k_{z}$.
To comply with the notation originally introduced in Ref.\cite{BTK-1982} to describe tunneling through a barrier in a superconductor, wave vectors with the plus (minus) sign inside the square root in Eq.(\ref{k-perp_vs_energy}) are referred to as electron-like (hole-like) wave vectors.

Depending on the value of $\varepsilon$ and the sign of $\tilde{\mu}$, there can be alternatively four complex solutions, four real solutions, and two real and two complex solutions of Eq.(\ref{k-perp_vs_energy}).
Only complex solutions resulting in decaying exponentials for $\rho \rightarrow \infty$ can be accepted.
A discussion of the explicit solutions in the various energy ranges, depending also on the sign of $\tilde{\mu}$, is reported in Appendix A, where the boundary conditions at $\rho = R_{\mathrm{out}}$ between the numerical solutions of 
Eqs.(\ref{cylindrical-BdG-equations}) for $\rho \le R_{\mathrm{out}}$ and the analytical solutions of Eqs.(\ref{modified-cylindrical-BdG-equations}) for $\rho \ge R_{\mathrm{out}}$ are also reported.

What is relevant here is that, depending on the allowed solutions $k_{\perp}$ to Eq.(\ref{k-perp_vs_energy}), the solutions
$(u_{\nu}(\rho),v_{\nu}(\rho))$ of the BdG equations for $\rho \ge R_{\mathrm{out}}$ can be expressed as linear combinations of
Bessel $J_{\alpha}(\zeta)$, Neumann $Y_{\alpha}(\zeta)$, and Hankel $H_{\alpha}^{\pm}(\zeta)$ functions of index $\alpha=|\ell'|$
and argument $\zeta=k_{\perp} \rho$.
These functions, in turn, have the following asymptotic behaviors (that holds for $\rho \gg R_{\mathrm{out}}$) \cite{AS-1972}:
\begin{eqnarray}
J_{\alpha}(\zeta) & \sim & \sqrt{\frac{2}{\pi \zeta}} \,\, \cos\left(\zeta - \frac{1}{2} \pi \alpha - \frac{1}{4} \pi \right)  \nonumber \\
Y_{\alpha}(\zeta) & \sim & \sqrt{\frac{2}{\pi \zeta}} \,\, \sin\left(\zeta - \frac{1}{2} \pi \alpha - \frac{1}{4} \pi \right)  
\label{asymptotic-behavior} \\
H_{\alpha}^{\pm}(\zeta) & \sim & \sqrt{\frac{2}{\pi \zeta}} \,\, \exp \left[ \pm \, i \left( \zeta - \frac{1}{2} \pi \alpha - \frac{1}{4} \pi \right) \right] \, .
\nonumber
\end{eqnarray}
\noindent
The  behaviors (\ref{asymptotic-behavior}) are what is only needed to calculate the normalization of the wave functions in the continuum part of the spectrum, to be considered next.

\vspace{0.05cm}
\begin{center}
{\bf C. Normalization in the continuum}
\end{center}
\vspace{0.05cm}

The normalization of the (two-component) wave functions, that are solutions of the BdG equations (\ref{cylindrical-BdG-equations}) 
for energy eigenvalues lying in the continuum, can be obtained by adapting to the present context the method discussed in 
Ref.\cite{GFS-1979} for the Schr\"{o}dinger equation.

Let us consider the BdG equations (\ref{cylindrical-BdG-equations}) for two different energies $\varepsilon$ and $\varepsilon'$, both lying
in the continuum. 
Multiplying these equations from the left by the pair $(u_{\varepsilon'},v_{\varepsilon'})$ and 
$(u_{\varepsilon},v_{\varepsilon})$, in the order, subtracting the resulting expressions side by side, and integrating over the radial coordinate from $\rho = 0$ up to $\rho = \bar{R}$, we obtain:
\begin{eqnarray}
& & \int_{0}^{\bar{R}} \! d\rho \, \rho \, \left[ u_{\varepsilon'}^{\lambda'}(\rho) \, u_{\varepsilon}^{\lambda}(\rho) \, + \, 
v_{\varepsilon'}^{\lambda'}(\rho) \, v_{\varepsilon}^{\lambda}(\rho) \right] 
\label{normalization_in_the_continuum}  \\
& & = \, \frac{\bar{R}}{2 m} \, \frac{\mathcal{P}}{(\varepsilon - \varepsilon')} \,  \left[ - u_{\varepsilon'}^{\lambda'}(\rho) \, 
\frac{d u_{\varepsilon}^{\lambda}(\rho)}{d \rho} \, + \, u_{\varepsilon}^{\lambda}(\rho) \, \frac{d u_{\varepsilon'}^{\lambda'}(\rho)}{d \rho} \right. 
\nonumber \\
& & \hspace{2.8cm} + \left. v_{\varepsilon'}^{\lambda'}(\rho) \, \frac{d v_{\varepsilon}^{\lambda}(\rho)}{d \rho} 
\, - \, v_{\varepsilon}^{\lambda}(\rho) \, \frac{d v_{\varepsilon'}^{\lambda'}(\rho)}{d \rho} \right]_{\rho=\bar{R}}
\nonumber
\end{eqnarray}
\noindent
where the index $\lambda$ distinguishes degenerate independent solutions (cf. Appendix A) and an integration by parts has been performed.
In the expression (\ref{normalization_in_the_continuum}), the two limits $\bar{R} \rightarrow \infty$ and $\varepsilon \rightarrow \varepsilon'$ 
have been taken in the order.
Note how the division by $(\varepsilon - \varepsilon')$ is interpreted as a principal part value ($\mathcal{P}$), consistently with the ``standing-wave boundary conditions'' we are adopting for the radial problem.
Note further that the (extreme) asymptotic form of the wave functions is what is only needed to establish their normalization.

In particular, for an asymptotic form of the type (with real values of $k_{\perp}$):
\begin{equation}
\left( \begin{array}{c} u_{\varepsilon}^{\lambda}(\rho) \\ v_{\varepsilon}^{\lambda}(\rho) \end{array} \right) \, = \,
\left( \begin{array}{c} u_{k} \\ v_{k} \end{array} \right) \,
\left[ c_{\lambda} J_{\alpha}(k_{\perp} \rho) \, + \,  d_{\lambda} Y_{\alpha}(k_{\perp} \rho) \right]    
\label{specific-asymptotic-form}
\end{equation}
\noindent
where $c_{\lambda}$ and $d_{\lambda}$ are real coefficients, in the appropriate limits the expression 
(\ref{normalization_in_the_continuum}) reduces to: 
\begin{eqnarray}
& & \int_{0}^{\infty} \! d\rho \, \rho \, \left[ u_{\varepsilon'}^{\lambda'}(\rho) \, u_{\varepsilon}^{\lambda}(\rho) \, + \, 
v_{\varepsilon'}^{\lambda'}(\rho) \, v_{\varepsilon}^{\lambda}(\rho) \right] 
\nonumber \\
& & = \left[ c_{\lambda} c_{\lambda'} \, + \,  d_{\lambda} d_{\lambda'} \right] \,\, \frac{1}{k_{\perp}} \,\, \delta(k_{\perp} - k_{\perp}') \, .
\label{specific-normalization_in_the_continuum} 
\end{eqnarray}
\noindent
To obtain this result we have made use of the identity:
\begin{equation}
\frac{1}{\varepsilon} \,  \frac{\mathcal{P}}{(\varepsilon - \varepsilon')} \, = \, \frac{m}{\left(\frac{k_{\perp}^{2}}{2m} - \tilde{\mu}\right)} \,
\frac{1}{k_{\perp}} \,  \frac{\mathcal{P}}{(k_{\perp} - k_{\perp}')}                           \label{relation_between_principal-parts}
\end{equation}
\noindent
that holds in the limit $\varepsilon \rightarrow \varepsilon'$.
A simple generalization of the expression (\ref{specific-normalization_in_the_continuum}) can be obtained when more than one wave vector appear on the right-hand side of Eq.(\ref{specific-asymptotic-form}).
\vspace{-0.3cm}

\section{III. Spatial profiles of a vortex from zero to the critical temperature}
\label{sec:spatial_profile}
\vspace{-0.3cm}

The solution of the BdG equations for an isolated vortex embedded in an infinite superfluid, discussed in Section II, enables us to obtain the spatial profile $\Delta(\rho)$ of the gap parameter (via the regularized gap equation (\ref{modified-gap-equation-II}) of Appendix B), as well as of the number $n(\rho)$ and current $j(\rho)$ densities (whose asymptotic contributions are given by Eqs.(\ref{asymptotic-density}) and (\ref{asymptotic-current-density}) of Appendix C, respectively).

In the following, the chemical potential $\mu$ entering the BdG equations is eliminated in favor of the asymptotic (bulk) value $n_{0}$ of the density via the standard BCS density equation for a homogeneous system in the absence 

\begin{figure}[h]
\includegraphics[angle=0,width=7.5cm]{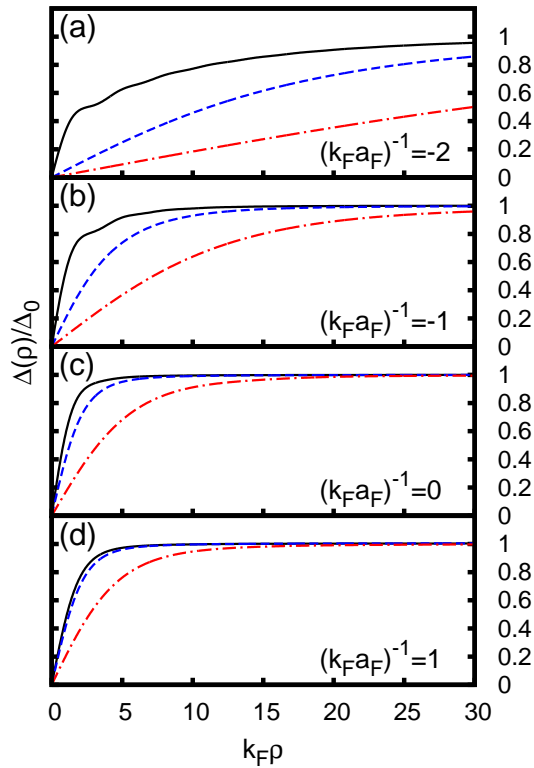}
\caption{Gap parameter $\Delta(\rho)$ (normalized to its asymptotic value $\Delta_{0}$ at the given temperature) of an isolated vortex  versus the distance $\rho$ from the center, for the coupling $(k_{F} a_{F})^{-1}$: (a) $-2.0$; (b) $-1.0$; (c) $0.0$; (d) $+1.0$. For each coupling, three different temperatures are considered: $T=0$ (full lines); $T=0.5T_{c}$ (dashed lines); $T=0.9T_{c}$ (dashed-dotted lines).}
\label{Figure-2}
\end{figure}

\noindent
of the vortex, namely, 
\begin{equation}
n_{0} = \int \! \frac{d\mathbf{k}}{(2 \pi)^{3}} 
\left[ 1 - \frac{\xi_{\mathbf{k}}}{E_{\mathbf{k}}} \left( 1 - 2 f_{F}(E_{\mathbf{k}}) \right) \right] 
\label{coarse-garined-density}
\end{equation}
\noindent
where $\xi_{\mathbf{k}} = \frac{\mathbf{k}^{2}}{2m} - \mu$ and $E_{\mathbf{k}} = \sqrt{\xi_{\mathbf{k}}^{2} + \Delta_{0}^{2}}$, since corrections to $\mu$ due to the presence of an isolated vortex are negligible in the thermodynamic limit.
This procedure, in turn, fixes the value of $k_{F}$.

\begin{figure}[h]
\includegraphics[angle=0,width=7.5cm]{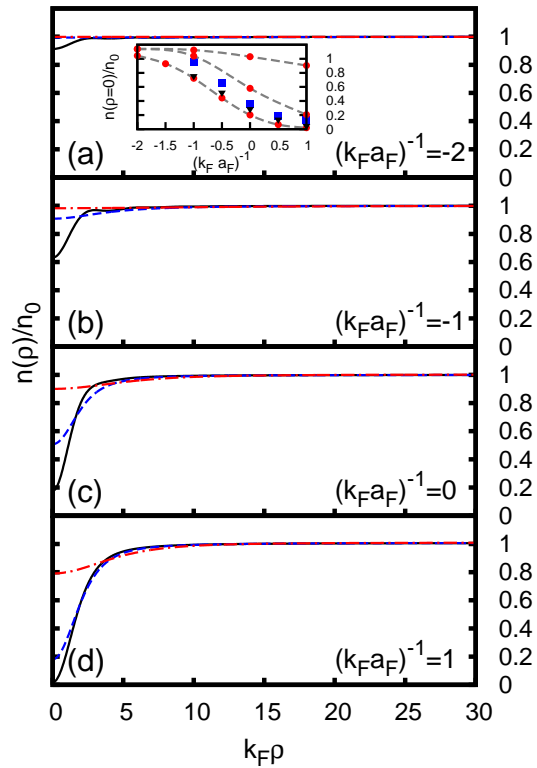}
\caption{Number density $n(\rho)$ (normalized to its asymptotic value $n_{0}$) for an isolated vortex versus the distance $\rho$ from the center, for the same couplings and temperatures of Fig.÷\ref{Figure-2}.
The inset shows the density at the center of the vortex vs $(k_{F} a_{F})^{-1}$, where our results at $T = (0,0.5,0.9) T_{c}$ from bottom to top (circles with interpolating dashed lines) are compared with those at $T=0$ from Ref.\cite{SRH-2006} (squares) and from Ref.\cite{Levin-2006} (triangles).}
\label{Figure-3}
\end{figure}

\begin{figure}[t]
\includegraphics[angle=0,width=7.5cm]{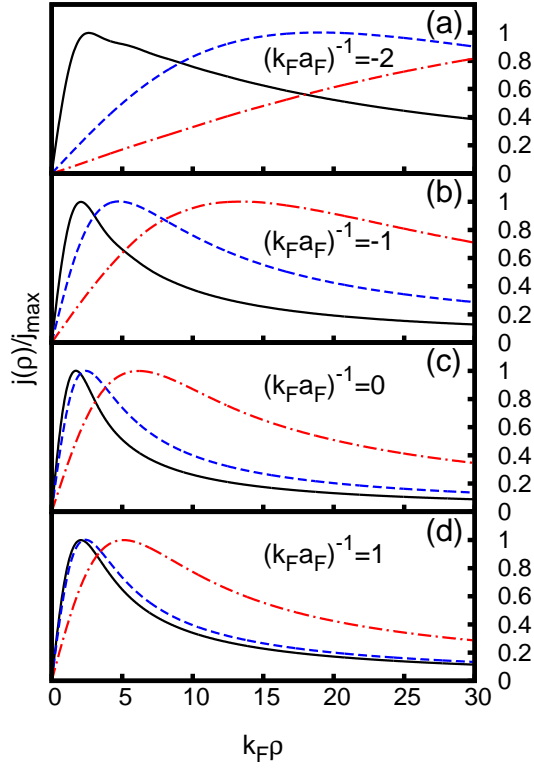}
\caption{Number current $j(\rho)$ (normalized to its maximum value $j_{\mathrm{max}}$ at the given temperature) of an isolated vortex versus the distance $\rho$ from the center, for the same couplings and temperatures of Figs.÷\ref{Figure-2} and \ref{Figure-3}. The maximum value of the current conventionally identifies the vortex radius $R_{\mathrm{v}}$.}
\label{Figure-4}
\end{figure}

Figures \ref{Figure-2}-\ref{Figure-4} show our numerical results for the quantities $\Delta(\rho)$, $n(\rho)$, and $j(\rho)$, in the order, for the four couplings $(k_{F} a_{F})^{-1} = (-2.0,-1.0,0.0,+1.0)$ and the three temperatures $T = (0.0,0.5,0.9)T_{c}$.
These plots were generated using a common cutoff energy $E_{c} = 3 E_{F}$ which, thanks to our novel regularization procedure (cf. Appendices B and C), proves sufficient to achieve maximum accuracy of the calculations to the extent that using larger values of $E_{c} $ provides essentially the same results.
A number of similar plots (not shown here) have also been systematically generated over a finer mesh of temperatures from $T=0$ up to $T=0.95T_{c}$, in order to extract from them the temperature dependence of the healing length associated with the vortex, as discussed in Section IV.

Note from Figs.÷\ref{Figure-2}-\ref{Figure-4} that the size of the vortex increases more rapidly with increasing temperature when approaching the BCS limit 
$(k_{F} a_{F})^{-1} \lesssim -1$.
Note also the presence of the characteristic Friedel's oscillations in all these quantities when this limit is approached at low temperature.
These oscillations, however, fade away rather quickly as the temperature is increased toward $T_{c}$.

\begin{figure}[t]
\includegraphics[angle=0,width=7.5cm]{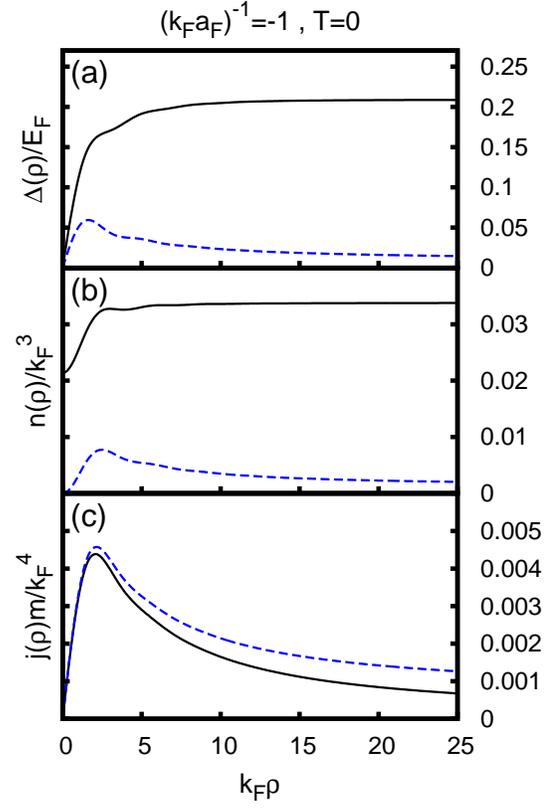}
\caption{(a) Gap parameter $\Delta(\rho)$, (b) number density $n(\rho)$, and (c) number current $j(\rho)$ of an isolated vortex versus the distance $\rho$ from the center, at zero temperature for the coupling $(k_{F} a_{F})^{-1} = -1$. 
The results of the complete calculation (full lines) are contrasted with those of a calculation that excludes the contribution from the continuum part of the spectrum in Eqs. (\ref{modified-gap-equation-II}), (\ref{BdG-density}), and (\ref{BdG-current}) (dashed lines).}
\label{Figure-5}
\end{figure}

As we have already mentioned, the reason why we have invested much effort in determining the continuum part of the spectrum of the BdG equations in an infinite medium is that this part is expected to exhaust in practice most part of the contribution to physical quantities.

In support to this expectation, we show in Fig.÷\ref{Figure-5} the profiles of $\Delta(\rho)$, $n(\rho)$, and $j(\rho)$ obtained at zero temperature for the coupling 
$(k_{F} a_{F})^{-1} = -1$, alternatively by including or omitting the contribution from the continuum part of the spectrum in the calculation of these quantities.
Drastic changes in these profiles result indeed when the contribution from the continuum is omitted from the calculation (with similar conclusions drawn for different temperatures and couplings).
A more complete analysis of how different energy ranges in the solutions of the BdG equations contribute to the spatial profiles of these physical quantities will be presented in Section V.
Note that an appropriate absolute normalization is used in Fig.÷\ref{Figure-5} for each quantity, in order to obtain a meaningful comparison.

\begin{figure}[h]
\includegraphics[angle=0,width=7.5cm]{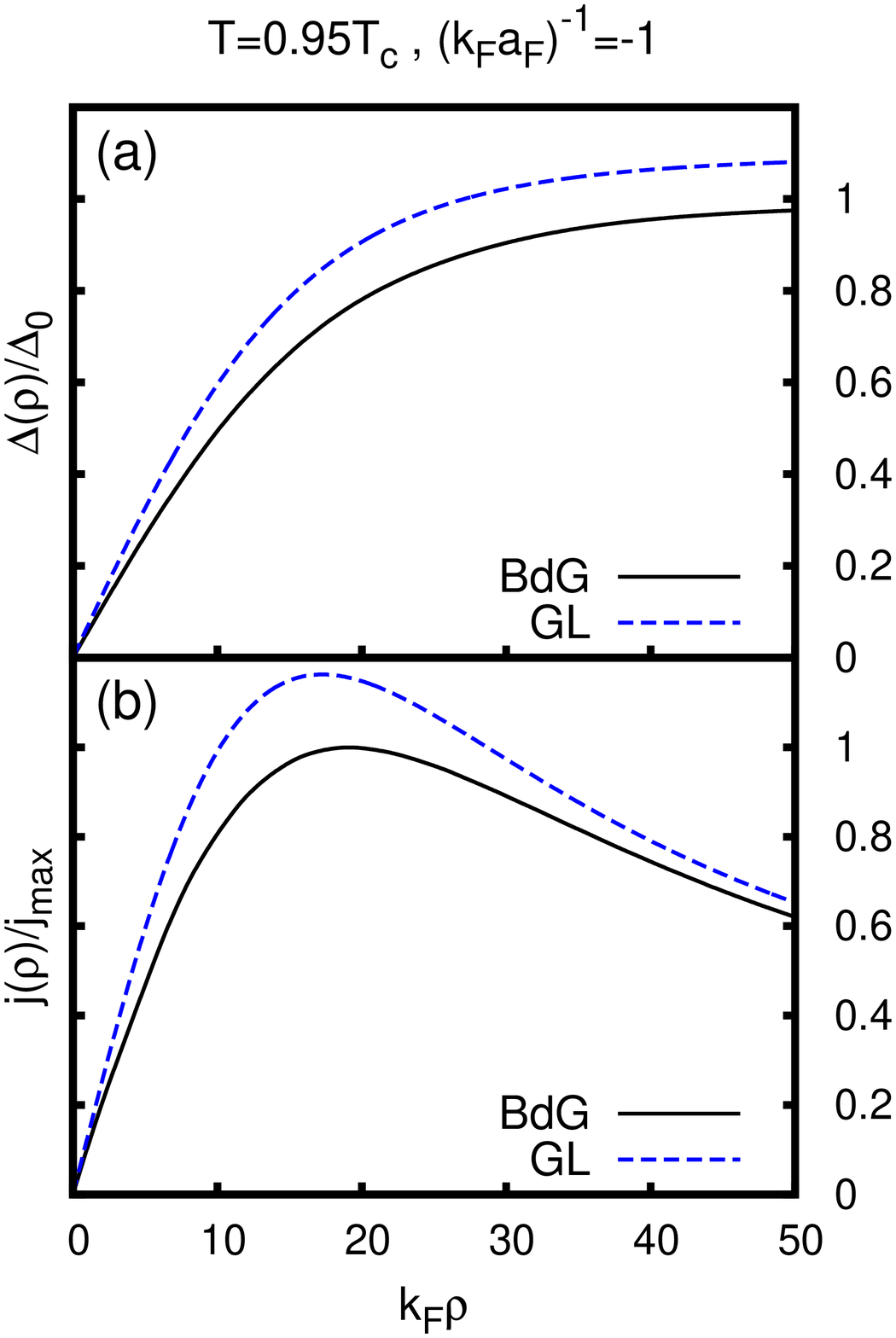}
\caption{(a) Gap parameter $\Delta(\rho)$ and (b) number current $j(\rho)$ of an isolated vortex versus the distance $\rho$ from the center, close to the critical temperature for the coupling $(k_{F} a_{F})^{-1} = -1$. The results of the calculation of the BdG equations (full lines) are compared with those obtained by the Ginzburg-Landau (GL) theory (dashed lines). The maximum values $\Delta_{0}$ for $\Delta(\rho)$ and $j_{\mathrm{max}}$ for $j(\rho)$ correspond to the BdG calculation.}
\label{Figure-6}
\end{figure}

\begin{figure}[t]
\includegraphics[angle=0,width=7.5cm]{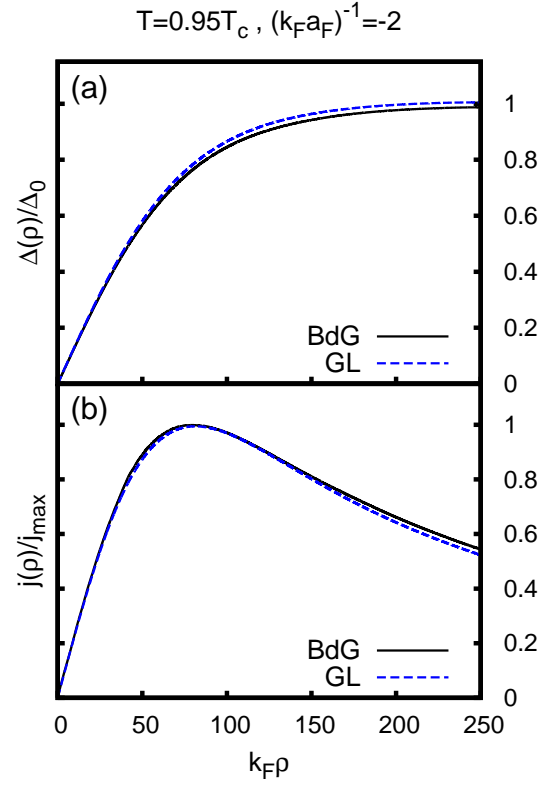}
\caption{Same as Fig.÷\ref{Figure-6} but for the coupling $(k_{F} a_{F})^{-1} = -2$ closer to the BCS limit.}
\label{Figure-7}
\end{figure}

It is further relevant to compare the profiles of the order parameter and the number current obtained by the present accurate solution of the BdG equations on the BCS side of the crossover close to $T_{c}$, with those obtained by the less demanding numerical solution of the Ginzburg-Landau (GL) differential equation for the order parameter $\Delta_{\mathrm{GL}}$, namely \cite{FW}:
\begin{eqnarray}
& & \left[ \frac{6 \pi^{2} (k_{B} T_{c})^{2} }{7 \zeta(3) E_{F}} \left(1 - \frac{T}{T_{c}} \right) + \frac{\nabla^{2}}{4 \, m}  \right] \, \Delta_{\mathrm{GL}}(\mathbf{r})
\nonumber \\
& & -  \frac{3}{4 E_{F}} \, |\Delta_{\mathrm{GL}}(\mathbf{r})|^{2} \Delta_{\mathrm{GL}}(\mathbf{r}) = 0
\label{GL-gap-equation}
\end{eqnarray}
\noindent
where $\zeta(3) \simeq 1.202$ is the Riemann zeta function of argument $3$.
In terms of this $\Delta_{\mathrm{GL}}(\mathbf{r})$, the GL current is then given by the expression \cite{FW}:
\begin{eqnarray}
\mathbf{j}_{\mathrm{GL}}(\mathbf{r}) & = & \frac{7 \, \zeta(3) \, n_{0}}{16 \, i \, m \, (\pi k_{B} T_{c})^{2}} 
\left[ \Delta_{\mathrm{GL}}(\mathbf{r})^{*} \, \nabla \, \Delta_{\mathrm{GL}}(\mathbf{r}) \right. 
\nonumber \\ 
& & \left. \hspace{2.3cm} - \, \Delta_{\mathrm{GL}}(\mathbf{r}) \, \nabla \, \Delta_{\mathrm{GL}}(\mathbf{r})^{*} \right] \, .
\label{GL-number-current}
\end{eqnarray}
Since the equation (\ref{GL-gap-equation}) for $\Delta_{\mathrm{GL}}$ and the expression (\ref{GL-number-current}) for $\mathbf{j}_{\mathrm{GL}}$ have been derived microscopically from the BdG equations in the (extreme) BCS limit \emph{and\/} close to the critical temperature \cite{Gorkov-1959}, one expects the numerical comparison with the full solution of the BdG equations to improve as these limiting conditions are approached.
That this is indeed the case is shown in Figs.÷\ref{Figure-6} and \ref{Figure-7}, where already for the coupling $(k_{F} a_{F})^{-1} = -2$ and the temperature
$T=0.95T_{c}$ the comparison between the two (BdG and GL) calculations appears quite good.

The above example can be regarded as a prototype for what was meant in the Introduction, about the fact that non-trivial numerical solutions of the BdG equations can be used in practice to test the validity of local equations for the order parameter under specific circumstances.

\vspace{-0.4cm}
\section{IV. Extracting the temperature dependence of the coherence length}
\label{sec:coherence_length}
\vspace{-0.2cm}

From the spatial profiles $\Delta(\rho)$ of the gap parameter for an isolated vortex that were obtained in Section III, we can now extract the characteristic coherence (healing) length as a function of temperature \emph{and} coupling according to the following procedure.

We note at the outset that, for given temperature, $\Delta(\rho)$ approaches its asymptotic (bulk) value $\Delta_{0}$ far away from the center of the vortex with the power-law behavior $\Delta_{0} (1 - \zeta^{2}/2\rho^{2})$, where $\zeta$ is a characteristic length.
In particular, in the BCS limit close to $T_{c}$, this behavior can be obtained directly from the GL equation (\ref{GL-gap-equation}) whereby $\zeta$ is identified with the GL coherence length \cite{FW}:
\begin{equation}
\xi_{\mathrm{GL}}(T) \, = \, \sqrt{\frac{7 \, \zeta(3) \, E_{F}}{24 \, m}} \, \frac{1}{\pi \, k_{B} T_{c}} \, \left( 1 \, - \, \frac{T}{T_{c}} \right)^{-1/2} \, .
\label{GL-coherence-length}
\end{equation}
\noindent
Similarly, in the BEC limit close to zero temperature, one can resort to the GP equation for composite bosons onto which the BdG equations map in that limit
\cite{PS-2003}, and identify $\zeta$ with the GP healing length $\xi_{GP}=(8 \pi a_{F} n_{0})^{-1/2}$.

Quite generally, for any coupling and temperature smaller than $T_{c}$, we have verified from the numerical solution of the BdG equations that $\Delta(\rho)$ always approaches its asymptotic value $\Delta_{0}$ like $\rho^{-2}$.
In practice, we have obtained the value of $\zeta$ through a fit of the type:

\begin{equation}
\Delta(\rho) = c_{0} \left( 1 \, - \, \frac{\zeta^{2}}{2 \, \rho^{2}} \right) \,\,\,\, \mathrm{when} \,\,\,\, 
\lambda R_{\mathrm{v}} \le \rho \lesssim (50 \div 150) k_{F}^{-1} 
\label{fitting-function-outer}
\end{equation}
\noindent
with $\lambda \sim 2 \div 6$ depending on coupling and temperature.
Here, $R_{\mathrm{v}}$ is the vortex radius identified from the profile of the current like in Fig.\ref{Figure-4}.
For smaller values of $\rho$, however, we have found that a separate exponential fit of the form
\begin{equation}
\Delta(\rho) = b_{0} \left( 1 \, - \, b_{1} \, e^{-\rho/\xi} \right) \,\, \mathrm{when} \,\,\, k_{F}^{-1} \le \rho \le \lambda R_{\mathrm{v}}
\label{fitting-function-inner}
\end{equation} 
\noindent
is more appropriate.
The need to exclude values of $k_{F} \rho$ smaller than one (at least on the BCS side of the crossover), in order to identify the length $\xi$ as in Eq.(\ref{fitting-function-inner}), was pointed out in Ref.\cite{SRH-2006} for an isolated vortex and in Ref.\cite{SPS-2010} in the context of the Josephson effect.

\begin{figure}[h]
\includegraphics[angle=0,width=7.5cm]{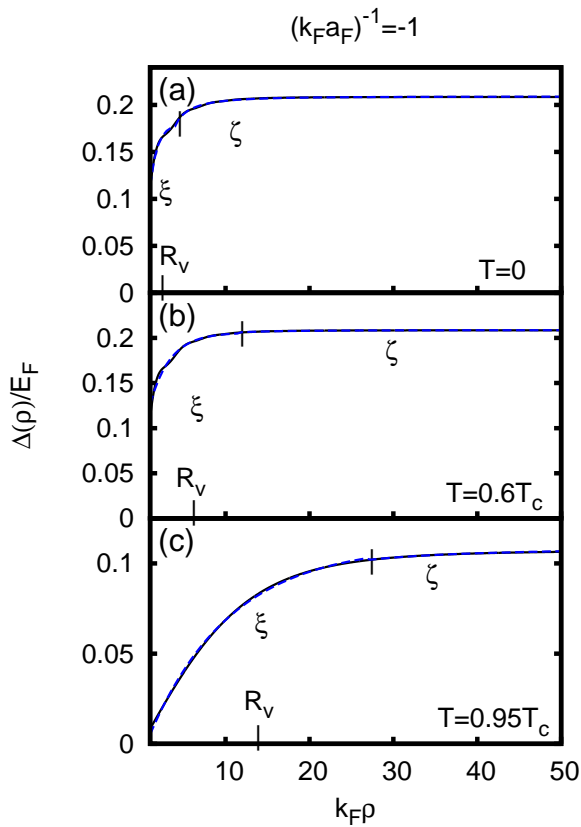}
\caption{Gap parameter $\Delta(\rho)$ (normalized to the Fermi energy $E_{F}$) of an isolated vortex versus the distance $\rho$ from the center, for three different temperature and $(k_{F} a_{F})^{-1} = -1$.
The results of the fittings to extract the lengths $\xi$ and $\zeta$ in the two different intervals of $\rho$ (broken lines) are compared with those of the the full calculation (full lines). The value of the vortex radius $R_{\mathrm{v}}$ is marked in each case.}
\label{Figure-8}
\end{figure}

The two independent fits (\ref{fitting-function-outer}) and (\ref{fitting-function-inner}) determine the two length scales $\zeta$ and $\xi$ which may, in principle, be different from each other.
We actually expect the ratio $\xi/\zeta$ not to be appreciably different from unity for \emph{all} couplings and temperatures, in such a way that \emph{a single length scale} can be eventually identified also from the BdG equations.
This would be similar to what occurs both in the BCS limit close to $T_{c}$ and in the BEC limit close to zero temperature, where a single length scale 
($\xi_{\mathrm{GL}}(T)$ and $\xi_{GP}$, in the order) is identified.

Figure ÷\ref{Figure-8} shows the typical quality of the fits (\ref{fitting-function-outer}) and (\ref{fitting-function-inner}) in the two adjacent spatial regions, for a specific coupling and three different temperatures.
These fits have then been repeated for several couplings about unitarity and for a rather dense mesh of temperatures.

\begin{figure}[h]
\includegraphics[angle=0,width=7.5cm]{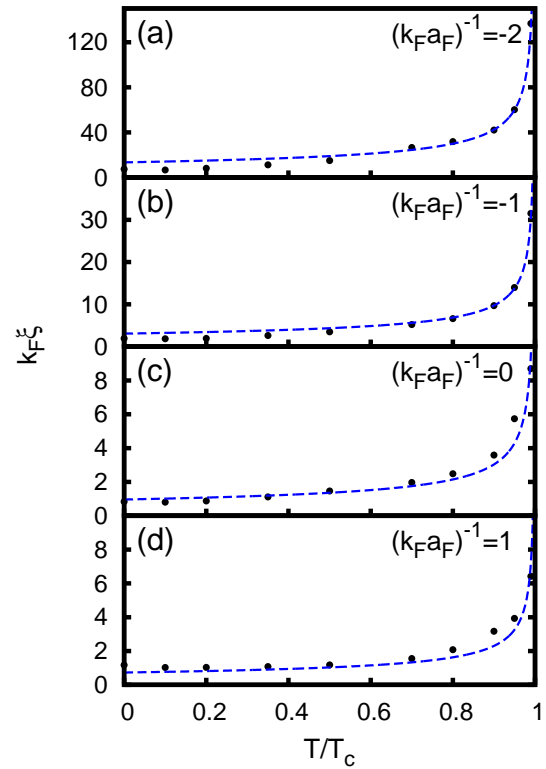}
\caption{The values of the healing length $\xi$, as obtained from the fits (\ref{fitting-function-inner}) to the profiles of $\Delta(\rho)$ (dots) for four different couplings across unitarity, are shown versus the temperature $T$ (in units of the respective critical temperature $T_{c}$).
These values are then fitted by the mean-field-like expression $\xi(T) \propto (T_{c} - T)^{-1/2}$ over the whole temperature range down to $T=0$ 
(dashed lines).}
\label{Figure-9}
\end{figure}

The values of the healing length $\xi$ extracted from these fits, for several couplings and from $T=0$ up to $T=0.99T_{c}$, are reported in Fig.÷\ref{Figure-9} as black dots.
Note again how $\xi$ increases faster with increasing temperature when the coupling progresses toward the BCS limit.
The dashed lines in Fig.÷\ref{Figure-9} are then obtained by assuming a simple expression of the form $k_{F} \xi(T) = A (1 - T/T_{c})^{-1/2}$ to hold for any coupling over the whole temperature range from $T=0$ up to (very close to) $T_{c}$, in analogy to the GL expression 
$k_{F} \xi_{\mathrm{GL}}(T) = A_{\mathrm{GL}} (1 - T/T_{c})^{-1/2}$ [for which $A_{\mathrm{GL}} = 0.47 E_{F}/\Delta_{0}(T=0)$, cf. Eq.(\ref{GL-coherence-length})] that holds in principle only in the (extreme) BCS limit quite close to $T_{c}$.
From this kind of fit we obtain the values $A = (13.41,3.08,0.96,0.73)$ for the four couplings $(k_{F} a_{F})^{-1} = (-2.0.-1.0,0.0,+1.0)$, in the order, which can be compared with the GL values $A_{\mathrm{GL}}  = (10.12,2.26,0.68)$ for the three couplings $(k_{F} a_{F})^{-1} = (-2.0.-1.0,0.0)$, values that are determined only in terms of the corresponding values of $\Delta_{0}(T=0)$.

\begin{figure}[t]
\includegraphics[angle=0,width=7.5cm]{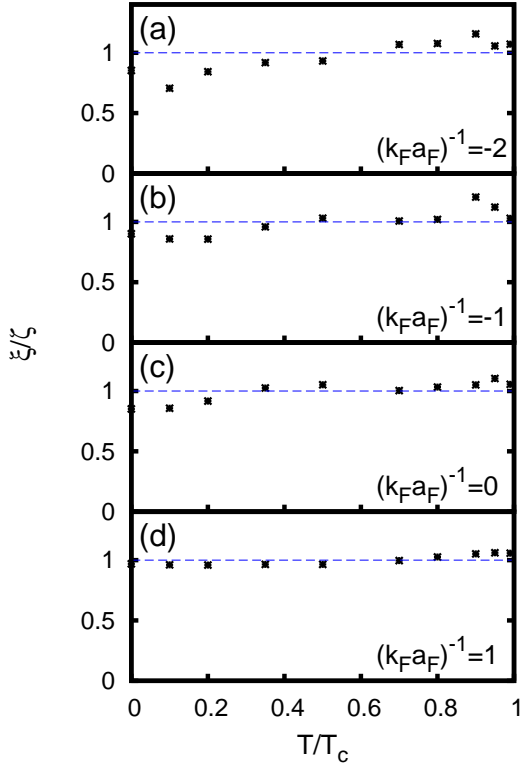}
\caption{Ratio of the two healing lengths $\xi$ and $\zeta$ (stars) for four different couplings across unitarity versus the temperature $T$ (in units of the respective critical temperature $T_{c}$). 
The horizontal (dashed) lines mark the value unity for $\xi/\zeta$.}
\label{Figure-10}
\end{figure}

Similar plots can be produced for the other length scale $\zeta$ extracted from the fits (\ref{fitting-function-outer}) to the profiles of $\Delta(\rho)$.
Figure ÷\ref{Figure-10} shows the ratio $\xi/\zeta$ between these two length scales as a function of temperature for several couplings.
It is rather remarkable that this ratio remains quite close to unity in all cases we have considered, thus justifying the statement that a \emph{single length scale} (say, the healing length $\xi$ of Eq.(\ref{fitting-function-inner})) can meaningfully be extracted from the BdG equations for all couplings and temperatures.
This conclusion will also be confirmed by a similar analysis about the temperature and coupling dependence of the vortex radius $R_{\mathrm{v}}$ reported in Appendix C.

\begin{figure}[t]
\includegraphics[angle=-90,width=7.5cm]{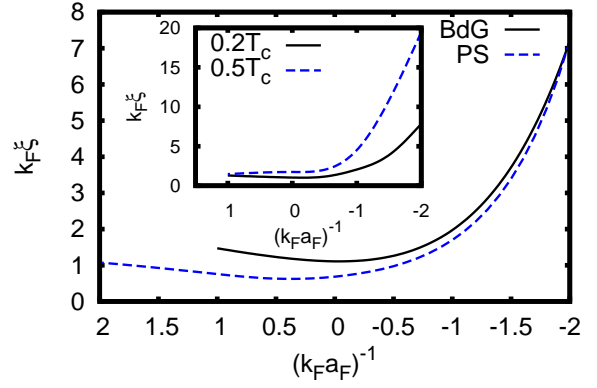}
\caption{Healing length $\xi$ at zero temperature versus the coupling $(k_{F} a_{F})^{-1}$, obtained from the present BdG calculation (full line) and from the approach of Ref.\cite{PS-1996} referred to as PS (dashed line).
For the sake of comparison, the two curves are made to coincide in the extreme BCS limit by multiplying the BdG curve by a factor 1.257, in order to account for their different definitions. 
In addition, the inset shows $\xi$ vs $(k_{F} a_{F})^{-1}$ obtained from the BdG calculation at the finite temperatures $T= 0.2 T_{c}$ and $T=0.5 T_{c}$ across the BCS-BEC crossover.}
\label{Figure-11}
\end{figure}

Finally, it is interesting to compare the values of the healing length $\xi$ at zero temperature across the BCS-BEC crossover, obtained by the present BdG analysis of the spatial profile of the gap parameter for an isolated vortex, with the alternative (and, in principle, unrelated) results for the so-called ``phase'' coherence length $\xi_{\mathrm{phase}}$, which were obtained in Ref.\cite{PS-1996} from the analysis of the spatial variation of the longitudinal component of the correlation function of the order parameter in an otherwise homogeneous system.
This comparison, presented in Fig.÷\ref{Figure-11}, shows a remarkable overall agreement between the coupling dependence of these two quantities, for which the minimum occurs at about unitarity in both cases.
In addition, the inset of Fig.÷\ref{Figure-11} presents similar curves obtained by the present BdG analysis at finite temperatures.
In this case, the minimum is seen to move for increasing $T$ progressively toward the BEC side of unitarity, as it is expected from the slower increase of the healing length $\xi$ for increasing temperature when the coupling progresses toward the BEC limit.

\vspace{-0.4cm}
\section{V. Contribution to the spatial profiles of physical quantities from different BdG energy ranges }
\label{sec:energy_contributions}
\vspace{-0.2cm}

We have already pointed out in Section III (see Fig.÷\ref{Figure-5} therein) that the continuum part of the spectrum of the BdG equations contributes in a substantial way to the spatial profiles of the gap parameter as well as of the number density and current.

In particular, we have obtained the result that the bound-state part of the spectrum which lies below the continuum threshold does not contribute to the density at the center of the vortex, so that in this case the contribution of the continuum part of the spectrum is overwhelming.
On the other hand, from the form of the analytic result (\ref{asymptotic-density}) for the asymptotic contribution to the density that originates from the continuum levels at high energy, one concludes that these levels, too, do not contribute to the density at the center of the vortex since $|\Delta(\mathbf{r})|$
vanishes therein.
It thus appears interesting to determine the way different energy ranges in the solutions of the BdG equations contribute to the spatial profiles of the above physical quantities. 

To this end, we introduce an ``upper limit'' $E_{\mathrm{ul}}$ for the energy such that only eigenstates of the BdG equations with 
$\varepsilon_{\nu} < E_{\mathrm{ul}} - \mu$ are retained in the calculation of the \emph{partial gap parameter} and of the \emph{partial number density and current}.
We then increase $E_{\mathrm{ul}}$ progressively starting from its value at the continuum threshold, which corresponds to 
$E_{\mathrm{ul}} - \mu = \Delta_{0}$ when $\mu >0$ and to $E_{\mathrm{ul}} - \mu = \sqrt{\Delta_{0}^{2} + \mu^{2}}$ when $\mu <0$,
reaching large values of $E_{\mathrm{ul}}$ to include eventually the high-energy part of the continuum.

\begin{figure}[h]
\includegraphics[angle=0,width=8.7cm]{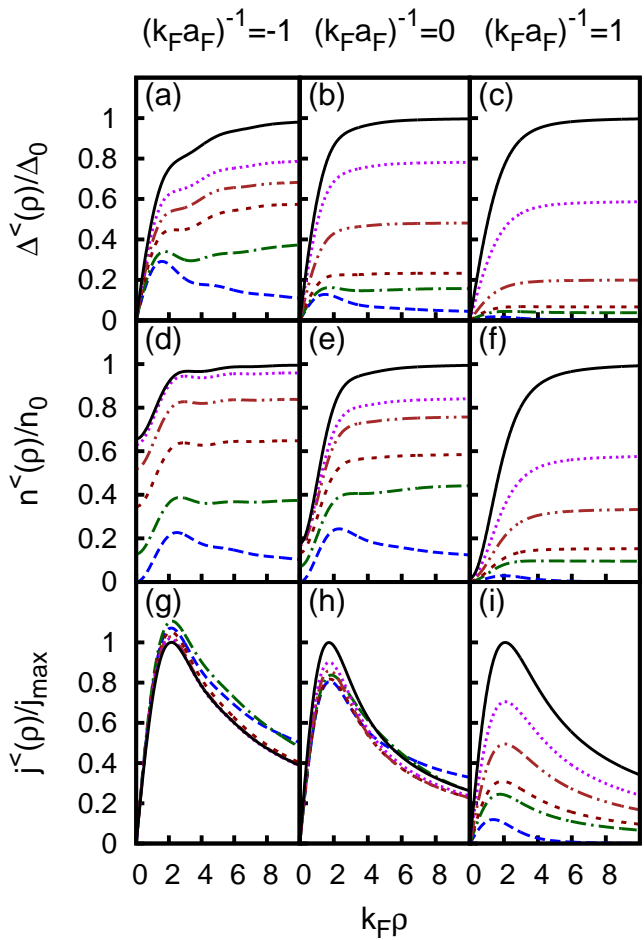}
\caption{Partial radial profiles of the gap parameter $\Delta^{<}(\rho)$, density $n^{<}(\rho)$, and current $j^{<}(\rho)$ at zero temperature for three couplings
$(k_{F} a_{F})^{-1}=(-1.0,0.0,+1.0)$, obtained using different values of the upper limit $E_{\mathrm{ul}}$ for the energy.
For convenience, the correspondence between the various values of $E_{ul}$ and the types of lines used in these plots for the three different couplings is reported separately in Table \ref{table-1} below.}
\label{Figure-12}
\end{figure}

Accordingly, the partial value $\Delta^{<}(\mathbf{r})$ of the gap parameter, that includes only eigenstates up to $E_{\mathrm{ul}} - \mu$, can be obtained from Eq.(\ref{modified-gap-equation-II}) of Appendix B by discarding the contribution of eigenstates with energy above $E_{\mathrm{ul}} - \mu$, thus writing 
in the place of Eq.(\ref{modified-gap-equation-II}):
\begin{eqnarray}
& & \left( - \frac{m}{4 \pi a_{F}} \, + \, \mathcal{R}(k_{c}) \right) \Delta^{<}(\mathbf{r}) \, =
\nonumber \\
& & = \sum_{\nu}^{\varepsilon_{\nu} < E_{ul} - \mu} u_{\nu}(\mathbf{r}) v_{\nu}(\mathbf{r})^{*}  
\left[ 1 - 2 f_{F}(\varepsilon_{\nu}) \right]
\label{partial-gap-equation}
\end{eqnarray}
\noindent
where $\mathcal{R}(k_{c})$ is defined by Eq.(\ref{definition-R(kc)}) of Appendix B.
Correspondingly, the partial values $n^{<}(\mathbf{r})$ for the density and $\mathbf{j}^{<}(\mathbf{r})$ for the current are obtained from the expressions (\ref{BdG-density}) and (\ref{BdG-current}) reported in Appendix C, where now the $\sum_{\nu}$ is limited to energies
$\varepsilon_{\nu} < E_{\mathrm{ul}} - \mu$.
For internal consistency, however, the eigenstates $u_{\nu}(\mathbf{r})$ and $v_{\nu}(\mathbf{r})$ utilized in these partial expressions are calculated from the BdG equations with the correct self-consistent value of $\Delta(\mathbf{r})$ which includes the contribution from all eigenstates.

\begin{table}[h]
\begin{tabular}{|c|c|c|c|}  \hline \hline
$(k_{F} a_{F})^{-1}$ & -1.0 & 0.0 & +1.0 \\  \hline \hline
long-dashed line      & $\,\,\, 1.16 \,\,\,$ & $\,\,\, 1.28 \,\,\,$ & $\,\,\, 0.75 \,\,\,$ \\  \hline
dotted-dashed line   & $\,\,\, 1.3 \,\,\,$ & $\,\,\, 1.4 \,\,\,$ & $\,\,\, 1.2 \,\,\,$ \\  \hline
short-dashed line     & $\,\,\, 1.5 \,\,\,$ & $\,\,\, 1.6 \,\,\,$ & $\,\,\, 1.5 \,\,\,$ \\  \hline 
double-dotted dashed line & $\,\,\, 1.7 \,\,\,$ & $\,\,\, 3.0 \,\,\,$ & $\,\,\, 2.0 \,\,\,$ \\  \hline 
dotted line & $\,\,\, 2.0 \,\,\,$ & $\,\,\, 6.0 \,\,\,$ & $\,\,\, 9.0 \,\,\,$ \\  \hline 
full line & $\,\,\, +\infty \,\,\,$ & $\,\,\, +\infty \,\,\,$ & $\,\,\, +\infty \,\,\,$ \\  \hline  \hline
\end{tabular}
\caption{Correspondence between the values of $E_{ul}$ (in units of $E_{F}$) and the types of lines used in Fig.÷\ref{Figure-12} for the three different couplings there considered. In all cases, the smallest value of $E_{ul}$ corresponds to the continuum threshold.}
\label{table-1}
\end{table}

The result of this calculation at zero temperature is reported in Fig.÷\ref{Figure-12} for three characteristic couplings.
Particularly striking appears here the result for the density for the couplings $-1.0$ and $0.0$ [cf. panels (d) and (e) of Fig.÷\ref{Figure-12}], for which the finite value at the center of the vortex is mostly contributed by continuum eigenstates quite close to threshold while no contribution is provided by the bound states below threshold.
In particular, when $(k_{F} a_{F})^{-1} = -1.0$ the value of $n^{<}(\rho = 0)$ passes from $20 \%$ to $99 \%$ of its full value when $E_{ul}$ varies from 
$1.30 E_{F}$ to $2.0 E_{F}$; and when $(k_{F} a_{F})^{-1} = 0.0$ from $40 \%$ to $95 \%$ of its full value when $E_{ul}$ varies from 
$1.40 E_{F}$ to $3.0 E_{F}$.

The above finding, that no contribution to $n^{<}(\rho = 0)$ originates from the bound states, is related to the fact that all bound states turn out to correspond to the second type of solutions (\ref{indicial-conditions-2}) with $\ell \le 0$, such that all $v_{\nu}^{(2)}(\rho)$ (and thus the density) vanish when $\rho = 0$.
At the same time, these bound states contribute in a coherent fashion to the anti-clock-wise circulation of the current (cf. Eq.(\ref{BdG-current}) of Appendix C), such that their contribution to the current may even exceed the value of the total current which includes also the contribution from the continuum  [cf. panels (g) and (h) of Fig.÷\ref{Figure-12}].
That the contribution of the (Andreev) bound states may sometimes exceed $100 \%$ of the total current was already pointed out in the context of the Josephson effect in Refs.\cite{Wendin-1996,SPS-2010}.

\vspace{-0.4cm}
\section{VI. Concluding remarks}
\label{sec:conclusions}
\vspace{-0.3cm}

In this paper, considerable efforts have been devoted to obtain an \emph{accurate} numerical solution of the fermionic BdG equations for a non-trivial but still manageable problem of an isolated vortex embedded in an otherwise infinite superfluid.
We have spanned the whole BCS-BEC crossover as a function of temperature up to $T_{c}$, in such a way that the spatial extension and the detailed shape of the vortex changes considerably as a function of both coupling and temperature.
To this end, we have left the vortex free of expanding out in the bulk of the superfluid not being constrained by walls, and implemented for the purpose the use of Òfree boundary conditionsÓ for the BdG equations.
We have also introduced a new regularization procedure for the gap equation that improves on previous proposals, so as to reduce the computational time while leaving unaltered the numerical accuracy. 

In this way, we have obtained the healing length for the vortex structure of the gap parameter as a function of both the temperature (from $T=0$ essentially up to $T=T_{c}$) and the coupling parameter $(k_{F} a_{F})^{-1}$.
This quantity shows an interesting behavior across the BCS-BEC crossover, which generalizes over the whole temperature vs coupling phase diagram what is already known from:
(i)  The GL approach in the weak-coupling (BCS) limit close to $T_{c}$;
(ii) The GP equation in the strong-coupling (BEC) limit at zero temperature;
(iii) The approach of Ref.\cite{PS-1996} across the BCS-BEC crossover at zero temperature.

In addition, by the present approach we have now at our disposal an accurate numerical solution of the BdG equations obtained for a non-trivial problem under a variety of circumstances, against which one might be able to compare the results of approximate differential equations that originate from local approximation of the BdG equations themselves.
These local (differential) equations could be, for instance, of the GL type in the weak-coupling (BCS) limit close to $T_{c}$ \cite{Gorkov-1959},
or of the GP type in the strong-coupling (BEC) limit at zero temperature \cite{PS-2003}.
In particular, still long awaited appears to be the comparison with the results obtained in the weak-coupling (BCS) limit away from $T_{c}$ deep in the superfluid phase, where generalizations of the GL equation have been attempted 
\cite{Tewordt-1963,Werthamer-1963,Werthamer-1964,Shanenko-2011} and deviations between the solutions of the BdG equations and these local equations are expected at low enough temperature.

The practical advantage of these differential equations stems from the fact that they are considerably simpler to solve than the original BdG equations, in such a way that, once their validity would have been explicitly tested against the results of the BdG equations in a number of manageable problems, they could be applied with confidence to the solution of more complex physical problems for which the use of the BdG equations remains prohibitive.
Work along these lines is in progress \cite{SPS-2012}.

\vspace{0.1cm}

\begin{center}
\begin{small}
{\bf ACKNOWLEDGMENTS}
\end{small}
\end{center}

We are grateful to A. Khan for his interest during a preliminary stage of this work and for having tested the regularization procedure for the gap equation at
zero temperature in a finite box.
This work was partially supported by the Italian MIUR under Contract Cofin-2009 ``Quantum gases beyond equilibrium''.
                                                                                                                                                                                                                                                                                                                                                                                                          
\appendix
\section{APPENDIX A: ENFORCING THE BOUNDARY CONDITIONS AT $R_{\mathrm{out}}$}
\label{sec:appendix_A}

In this Appendix, we describe in detail the solutions of the form (\ref{smart-solution}) that hold for $\rho \ge R_{\mathrm{out}}$ and are associated with the alternative values of $k_{\perp}$ obtained from Eq.(\ref{k-perp_vs_energy}), depending on the value of the energy $\varepsilon$ and the sign of 
$\tilde{\mu}$.
The solutions determined in this way for $\rho \ge R_{\mathrm{out}}$ will then be used to specify completely the wave functions obtained numerically for $\rho \le R_{\mathrm{out}}$, by enforcing the appropriate boundary conditions at $\rho = R_{\mathrm{out}}$. 

The method we use here is similar to that discussed in Ref.\cite{SPS-2010} for a one-dimensional geometry appropriate for the study of the Josephson effect throughout the BCS-BEC crossover, which extends the original approach of Ref.\cite{BTK-1982} that was limited to the (extreme) BCS limit.
A related approach was also used in Ref.\cite{BSTF-1987} for a gap parameter with spherical symmetry, and then utilized in 
Ref.\cite{YB-2003} to obtain the profile of a single vortex without enclosing it in a cylinder, again in the (extreme) BCS limit with a large coherence length at zero temperature.

When the positive energy $\varepsilon$ is increased from zero past the value $\sqrt{ \tilde{\mu}^{2} + \Delta_{0}^{2} }$, the four solutions for
$k_{\perp}$ given by Eq.(\ref{k-perp_vs_energy}) move in the complex $k_{\perp}$-plane. 
To follow their evolution versus $\varepsilon$, it is convenient to label these four solutions separately by adopting the convention:
\begin{eqnarray}
k_{\perp}^{(1)} & = & + \, \sqrt{2m \tilde{\mu}  \, + \, 2m \, \sqrt{ \varepsilon^{2} - \Delta_{0}^{2}} }    \nonumber \\ 
k_{\perp}^{(2)} & = &  - \, \sqrt{2m \tilde{\mu}  \, + \, \, 2m \, \sqrt{ \varepsilon^{2} - \Delta_{0}^{2}} }    \nonumber \\
k_{\perp}^{(3)} & = & + \, \sqrt{2m \tilde{\mu}  \, - \, \, 2m \, \sqrt{ \varepsilon^{2} - \Delta_{0}^{2}} }     \nonumber \\
k_{\perp}^{(4)} & = & - \, \sqrt{2m \tilde{\mu}  \, - \, \, 2m \, \sqrt{ \varepsilon^{2} - \Delta_{0}^{2}} } \, .    \label{the-four-k-perp_vs_energy}
\end{eqnarray}
\noindent
For $\rho \ge R_{\mathrm{out}}$, these wave vectors enter the arguments of the Bessel $J_{\alpha}(k_{\perp} \rho)$, Neumann 
$Y_{\alpha}(k_{\perp} \rho)$, and Hankel functions $H^{+}_{\alpha}(k_{\perp} \rho)$ functions (where $\alpha = |\ell'|$) with asymptotic behavior (\ref{asymptotic-behavior}), according to the following scheme.

\begin{figure}[t]
\includegraphics[angle=0,width=7.5cm]{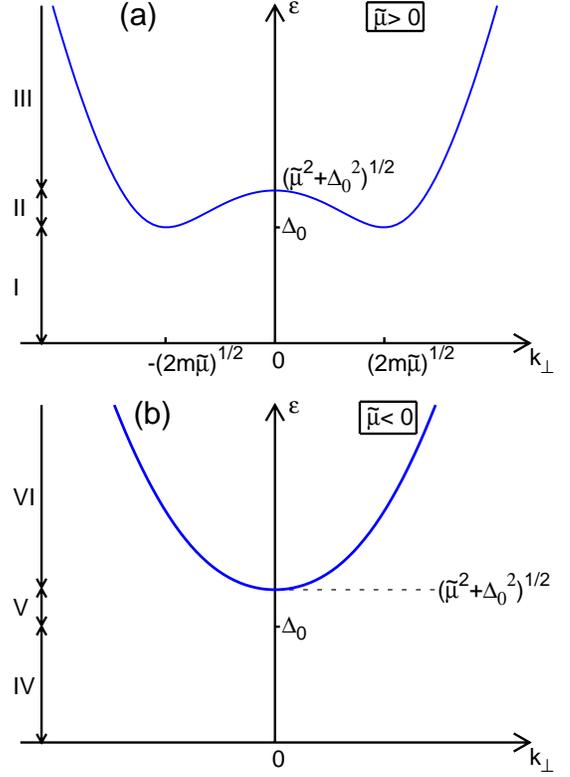}
\caption{Dispersion relation $\varepsilon$ vs $k_{\perp}$ given by Eq.(\ref{energy-eigenvalue}), with (a) $\tilde{\mu} > 0$ and (b) $\tilde{\mu} < 0$.
These plots identify the six energy ranges I-VI, where proper selections of the wave vectors (\ref{the-four-k-perp_vs_energy}) have alternatively to be done.}
\label{Figure-13}
\end{figure}

For the \emph{six ranges} that can be identified depending also on the sign of $\tilde{\mu}$ (as shown in Figs.÷\ref{Figure-13}(a) and \ref{Figure-13}(b), in the order), we obtain by inspection of the expressions (\ref{the-four-k-perp_vs_energy}):

\vspace{0.2cm}
\noindent
\emph{Range I}: $\tilde{\mu} > 0$ and $0 < \varepsilon < \Delta_{0}$.

\noindent
The $k_{\perp}^{(i)}$ ($i=1,\cdots,4$) are all complex, but only $k_{\perp}^{(1)}$ and $k_{\perp}^{(4)}$ have a positive imaginary part.
We then take alternatively $k_{\perp}^{(1)}$ and $k_{\perp}^{(4)}$ in the function $H^{+}_{\alpha}(k_{\perp} \rho)$.

\vspace{0.2cm}
\noindent
\emph{Range II}: $\tilde{\mu} > 0$ and  $\Delta_{0} < \varepsilon < \sqrt{ \tilde{\mu}^{2} + \Delta_{0}^{2} }$.

\noindent
In this case, $k_{\perp}^{(i)}$ ($i=1,\cdots,4$) are all real, and we take $k_{\perp}^{(1)}$ and $k_{\perp}^{(3)}$ in both functions 
$J_{\alpha}(k_{\perp} \rho)$ and $Y_{\alpha}(k_{\perp} \rho)$.

\vspace{0.2cm}
\noindent
\emph{Range III}: $\tilde{\mu} > 0$ and  $\varepsilon > \sqrt{ \tilde{\mu}^{2} + \Delta_{0}^{2} }$. 

\noindent
Here, $k_{\perp}^{(1)}$ and $k_{\perp}^{(2)}$ are real, and $k_{\perp}^{(3)}$ and $k_{\perp}^{(4)}$ are purely imaginary with 
Im$\{k_{\perp}^{(3)}\}>0$.
We thus take $k_{\perp}^{(1)}$ in both functions $J_{\alpha}(k_{\perp} \rho)$ and $Y_{\alpha}(k_{\perp} \rho)$, and $k_{\perp}^{(3)}$ in the function $H^{+}_{\alpha}(k_{\perp} \rho)$.

\vspace{0.2cm}
\noindent
\emph{Range IV}: $\tilde{\mu} < 0$ and $0 < \varepsilon < \Delta_{0}$.

\noindent
Same as for range I. We thus take alternatively $k_{\perp}^{(1)}$ and $k_{\perp}^{(4)}$ in the function $H^{+}_{\alpha}(k_{\perp} \rho)$.

\vspace{0.2cm}
\noindent
\emph{Range V}: $\tilde{\mu} < 0$ and  $\Delta_{0} < \varepsilon < \sqrt{ \tilde{\mu}^{2} + \Delta_{0}^{2} }$.

\noindent
In this case, all the $k_{\perp}^{(i)}$ ($i=1,\cdots,4$) are purely imaginary but only Im$\{k_{\perp}^{(1)}\}$ and Im$\{k_{\perp}^{(4)}\}$ are positive.
Again, we take alternatively $k_{\perp}^{(1)}$ and $k_{\perp}^{(4)}$ in the function $H^{+}_{\alpha}(k_{\perp} \rho)$.

\vspace{0.2cm}
\noindent
\emph{Range VI}: $\tilde{\mu} < 0$ and  $\varepsilon > \sqrt{ \tilde{\mu}^{2} + \Delta_{0}^{2} }$. 

\noindent
Here, $k_{\perp}^{(1)}$ and $k_{\perp}^{(2)}$ are real and $k_{\perp}^{(3)}$ and $k_{\perp}^{(4)}$ are purely imaginary, but only 
Im$\{k_{\perp}^{(4)}\}$ is positive.
Accordingly, we take $k_{\perp}^{(1)}$ in both functions $J_{\alpha}(k_{\perp} \rho)$ and $Y_{\alpha}(k_{\perp} \rho)$, and $k_{\perp}^{(4)}$ in the function $H^{+}_{\alpha}(k_{\perp} \rho)$.

\vspace{0.2cm}
With these premises, we pass now to enforce the \emph{boundary conditions} at $\rho = R_{\mathrm{out}}$, between the numerical solutions of the BdG equations (\ref{cylindrical-BdG-equations}) for $\rho \le R_{\mathrm{out}}$ discussed in subsection II-A and the analytic solutions of the modified BdG equations (\ref{modified-cylindrical-BdG-equations}) for $\rho \ge R_{\mathrm{out}}$ introduced in subsection II-B.

Ranges I, IV, and V as specified above can be dealt with in the same way, by writing the boundary conditions in the form (in the following equations, 
by $\ell'$ we shall actually mean its absolute value $|\ell'|$):
\begin{small}
\begin{eqnarray}
& & a \left( \begin{array}{c} u_{\varepsilon}^{(1)}(R_{\mathrm{out}}) \\ v_{\varepsilon}^{(1)}(R_{\mathrm{out}}) \end{array} \right) + \, 
b \left( \begin{array}{c} u_{\varepsilon}^{(2)}(R_{\mathrm{out}}) \\ v_{\varepsilon}^{(2)}(R_{\mathrm{out}}) \end{array} \right)  
\label{boundary-conditions-ranges1-4-5-for-functions} \\
& = & c \left( \begin{array}{c} u_{k_{1}} \\ v_{k_{1}} \end{array} \right) \, H^{+}_{\ell'}(k_{\perp}^{(1)} R_{\mathrm{out}}) \, + \, 
d \left( \begin{array}{c} u_{k_{4}} \\ v_{k_{4}} \end{array} \right) \, H^{+}_{\ell'}(k_{\perp}^{(4)} R_{\mathrm{out}})
\nonumber
\end{eqnarray}
\end{small}
\noindent
for the functions, and
\begin{small}
\begin{eqnarray}
& & a \left( \begin{array}{c} \frac{d u_{\varepsilon}^{(1)}(\rho)}{d \rho} \\ \frac{ d v_{\varepsilon}^{(1)}(\rho)}{d \rho} \end{array} \right)_{\rho=R_{\mathrm{out}}} + b \left( \begin{array}{c} \frac{d u_{\varepsilon}^{(2)}(\rho)}{d \rho} \\ \frac{ d v_{\varepsilon}^{(2)}(\rho)}{d \rho} \end{array} \right)_{\rho=R_{\mathrm{out}}}  
\label{boundary-conditions-ranges1-4-5-for-first-derivatives} \\
& = & c \left( \begin{array}{c} u_{k_{1}} \\ v_{k_{1}} \end{array} \right) \left.\frac{d H^{+}_{\ell'}(k_{\perp}^{(1)} \rho)}{d \rho} \right|_{R_{\mathrm{out}}}  
+ d \left( \begin{array}{c} u_{k_{4}} \\ v_{k_{4}} \end{array} \right) \left.\frac{d H^{+}_{\ell'}(k_{\perp}^{(4)} \rho)}{d \rho} \right|_{R_{\mathrm{out}}} 
\nonumber
\end{eqnarray}
\end{small}
\noindent
for their first derivatives.
Here, $(u_{\varepsilon}^{(1)}(\rho),v_{\varepsilon}^{(1)}(\rho))$ and $(u_{\varepsilon}^{(2)}(\rho),v_{\varepsilon}^{(2)}(\rho))$ are the two independent solutions of the BdG equations (\ref{cylindrical-BdG-equations}) identified by the indicial conditions (\ref{indicial-conditions-1}) and (\ref{indicial-conditions-2}), in the order.
The conditions (\ref{boundary-conditions-ranges1-4-5-for-functions}) and (\ref{boundary-conditions-ranges1-4-5-for-first-derivatives}) thus provide an algebraic homogeneous system of four equations in the four unknowns ($a,b,c,d$), which admits nontrivial solutions only for special values of $\varepsilon$, which correspond to the Andreev-Saint-James bound states associated with the spatial depression of the gap $\Delta(\rho)$ about $\rho = 0$.
In this case, the normalization of the single wave function $(u_{\varepsilon}(\rho),v_{\varepsilon}(\rho))$, as obtained by the linear combination on the left-hand side of Eq.(\ref{boundary-conditions-ranges1-4-5-for-functions}) for $\rho \le R_{\mathrm{out}}$ and on the right-hand side of Eq.(\ref{boundary-conditions-ranges1-4-5-for-functions}) for $\rho \ge R_{\mathrm{out}}$, is determined by:
\begin{equation}
 \int_{0}^{\infty} \! d\rho \, \rho \, \left[ u_{\varepsilon}(\rho) \, u_{\varepsilon}(\rho) \, + \, v_{\varepsilon}(\rho) \, v_{\varepsilon}(\rho) \right] \, = \, 1 \, .
\label{normalization-Andreev-Saint-James-bound-states}
\end{equation} 

Ranges III and VI can as well be treated on the same footing, by writing the boundary conditions in the form:
\begin{small}
\begin{eqnarray}
& & a \left( \begin{array}{c} u_{\varepsilon}^{(1)}(R_{\mathrm{out}}) \\ v_{\varepsilon}^{(1)}(R_{\mathrm{out}}) \end{array} \right) + \, 
b \left( \begin{array}{c} u_{\varepsilon}^{(2)}(R_{\mathrm{out}}) \\ v_{\varepsilon}^{(2)}(R_{\mathrm{out}}) \end{array} \right)  
\nonumber \\
& = & \left( \begin{array}{c} u_{k_{1}} \\ v_{k_{1}} \end{array} \right) \, 
\left[ c \, J_{\ell'}(k_{\perp}^{(1)} R_{\mathrm{out}}) \, + \,  d \, Y_{\ell'}(k_{\perp}^{(1)} R_{\mathrm{out}}) \right] 
\nonumber  \\
& + & e \left( \begin{array}{c} u_{k_{i}} \\ v_{k_{i}} \end{array} \right) \,  H^{+}_{\ell'}(k_{\perp}^{(i)} R_{\mathrm{out}}) 
\label{boundary-conditions-ranges3-6-for-functions} 
\end{eqnarray}
\end{small}
\noindent
and
\begin{small}
\begin{eqnarray}
& & a \left( \begin{array}{c} \frac{d u_{\varepsilon}^{(1)}(\rho)}{d \rho} \\ \frac{d v_{\varepsilon}^{(1)}(\rho)}{d \rho} \end{array} \right)_{\rho=R_{\mathrm{out}}} 
+ \, b \left( \begin{array}{c} \frac{d u_{\varepsilon}^{(2)}(\rho)}{d \rho} \\ \frac{d v_{\varepsilon}^{(2)}(\rho)}{d \rho} \end{array} \right)_{\rho=R_{\mathrm{out}}}  
\nonumber \\
& = & \left( \begin{array}{c} u_{k_{1}} \\ v_{k_{1}} \end{array} \right) \, 
\left[ c \, \left. \frac{d J_{\ell'}(k_{\perp}^{(1)} \rho)}{d \rho} \right|_{R_{\mathrm{out}}} \, + \,  d \, \left. \frac{d Y_{\ell'}(k_{\perp}^{(1)}  \rho)}{d \rho} \right|_{R_{\mathrm{out}}} \right] 
\nonumber \\
& + & e \left( \begin{array}{c} u_{k_{i}} \\ v_{k_{i}} \end{array} \right) \, \left. \frac{d H^{+}_{\ell'}(k_{\perp}^{(i)}  \rho)}{d \rho} \right|_{R_{\mathrm{out}}} 
\label{boundary-conditions-ranges3-6-for-first-derivatives} 
\end{eqnarray}
\end{small}
\noindent
where $i=3$ in range III and $i=4$ in range VI. 
The conditions (\ref{boundary-conditions-ranges3-6-for-functions}) and (\ref{boundary-conditions-ranges3-6-for-first-derivatives}) correspond to four algebraic equations in the five unknowns ($a,b,c,d,e$).
The normalization condition (\ref{specific-normalization_in_the_continuum}) with $\lambda=\lambda'$ for $\varepsilon$ in the continuum then provides a fifth condition for the coefficients
$c$ and $d$, that permits to determine all coefficients uniquely.

Finally, range II requires a slightly different handling because the electron-like and hole-like wave vectors are both real.
We then apply the boundary conditions to the functions $(u_{\varepsilon}^{(1)}(\rho),v_{\varepsilon}^{(1)}(\rho))$ and 
$(u_{\varepsilon}^{(2)}(\rho),v_{\varepsilon}^{(2)}(\rho))$ separately and write:
\begin{small}
\begin{eqnarray}
& & \left( \begin{array}{c} u_{\varepsilon}^{(1)}(R_{\mathrm{out}}) \\ v_{\varepsilon}^{(1)}(R_{\mathrm{out}}) \end{array} \right) 
\label{boundary-conditions-range2-for-functions}   \\
& = & \left( \begin{array}{c} u_{k_{1}} \\ v_{k_{1}} \end{array} \right) \, 
\left[ c_{11} \, J_{\ell'}(k_{\perp}^{(1)} R_{\mathrm{out}}) \, + \,  d_{11} \, Y_{\ell'}(k_{\perp}^{(1)} R_{\mathrm{out}}) \right] 
\nonumber \\
& + & \left( \begin{array}{c} u_{k_{3}} \\ v_{k_{3}} \end{array} \right) \, 
\left[ c_{13} \, J_{\ell'}(k_{\perp}^{(3)} R_{\mathrm{out}}) \, + \,  d_{13} \, Y_{\ell'}(k_{\perp}^{(3)} R_{\mathrm{out}}) \right] 
\nonumber
\end{eqnarray}
\end{small}
\noindent
and
\begin{footnotesize}
\begin{eqnarray}
& & \left( \begin{array}{c} \frac{d u_{\varepsilon}^{(1)}(\rho)}{d\rho} \\ \frac{dv_{\varepsilon}^{(1)}(\rho)}{d\rho} \end{array} \right)_{\rho=R_{\mathrm{out}}} 
\label{boundary-conditions-range2-for-first-derivatives} \\
& = & \left( \begin{array}{c} u_{k_{1}} \\ v_{k_{1}} \end{array} \right) \, 
\left[ c_{11} \, \left. \frac{d J_{\ell'}(k_{\perp}^{(1)} \rho)}{d\rho} \right|_{\rho=R_{\mathrm{out}}} \, + \,  d_{11} \, \left. \frac{d Y_{\ell'}(k_{\perp}^{(1)} \rho)}{d\rho}  \right|_{\rho=R_{\mathrm{out}}}\right] 
\nonumber \\
& + & \left( \begin{array}{c} u_{k_{3}} \\ v_{k_{3}} \end{array} \right) \, 
\left[ c_{13} \, \left. \frac{d J_{\ell'}(k_{\perp}^{(3)} \rho)}{d\rho}  \right|_{\rho=R_{\mathrm{out}}} \, + \,  d_{13} \, \left. \frac{d Y_{\ell'}(k_{\perp}^{(3)} \rho)}{d\rho}  \right|_{\rho=R_{\mathrm{out}}}\right] \, .
\nonumber
\end{eqnarray}
\end{footnotesize}
\noindent
Here, the four coefficients ($c_{11},d_{11},c_{13},d_{13}$) can be uniquely determined in terms of the known constants given by the left-hand side of
Eqs.(\ref{boundary-conditions-range2-for-functions}) and (\ref{boundary-conditions-range2-for-first-derivatives}).
Similar conditions are obtained for the second function $(u_{\varepsilon}^{(2)}(\rho),v_{\varepsilon}^{(2)}(\rho))$.
However, the two functions $(u_{\varepsilon}(\rho),v_{\varepsilon}(\rho))$ obtained in this way for all values of $\rho$ are not properly normalized in the continuum and are not orthogonal to each other.
In this case, the orthonormalization condition (\ref{normalization_in_the_continuum}) reads:
\begin{eqnarray}
& & \int_{0}^{\infty} \! d\rho \, \rho \, \left[ u_{\varepsilon'}^{\lambda'}(\rho) \, u_{\varepsilon}^{\lambda}(\rho) \, + \, 
v_{\varepsilon'}^{\lambda'}(\rho) \, v_{\varepsilon}^{\lambda}(\rho) \right] 
\label{ortho-normalization_in_the_continuum-for-range2} \\
& & = \left[ c_{\lambda 1} c_{\lambda' 1} \, + \,  d_{\lambda 1} d_{\lambda' 1} \right] \,\, \frac{1}{ k_{\perp}^{(1)} } \,\, \delta( k_{\perp}^{(1)} - k_{\perp}^{(1)'} ) 
\nonumber \\
& & + \left[ c_{\lambda 3} c_{\lambda' 3} \, + \,  d_{\lambda 3} d_{\lambda' 3} \right] \,\, \frac{1}{ k_{\perp}^{(3)} } \,\, \delta( k_{\perp}^{(3)} - k_{\perp}^{(3)'} )
\nonumber 
\end{eqnarray}
\noindent
where $\lambda,\lambda'=(1,2)$.

\appendix
\section{APPENDIX B: REGULARIZATION PROCEDURE FOR THE SELF-CONSISTENT GAP EQUATION}
\label{sec:appendix_B}

It is well known that the self-consistent condition (\ref{self-consistency}) for the gap parameter $\Delta(\mathbf{r})$ diverges in the ultraviolet in the case
of a contact inter-particle potential with coupling constant $v_{0}$ and has to be regularized accordingly.

In the homogeneous case with a uniform gap parameter $\Delta_{0}$, this regularization is readily achieved by expressing the bare coupling constant $v_{0}$ that enters Eq.(\ref{self-consistency}) in terms of the scattering length $a_{F}$ of the two-body problem, via the relation
\begin{equation}
- \,  \frac{1}{v_{0}} \, = \, - \, \frac{m}{4 \, \pi \, a_{F}} \, + \, \int^{k_{0}} \!\!\! \frac{d\mathbf{k}}{(2 \pi)^{3}} \,\, \frac{m}{\mathbf{k}^{2}} \, .
\label{homogeneous-regularization}
\end{equation}
\noindent
Here, $k_{0}$ is an ultraviolet cutoff which is eventually let $\rightarrow \infty$ while $v_{0} \rightarrow 0$ by keeping $a_{F}$ at the desired value.

This simple regularization, however, cannot be exploited when the gap parameter $\Delta(\mathbf{r})$ has a spatial dependence occurring, for instance, in the presence of an isolated vortex as considered in the present paper, or, more generally, in the presence of a scalar trapping potential $V_{\mathrm{ext}}(\mathbf{r})$ or of an effective vector potential $\mathbf{A}(\mathbf{r})$, the latter arising when the trap is set into rotation \cite{Ketterle-2005} or artificial gauge potentials are applied to neutral atoms \cite{Dalibard-2011}.
In all these cases, a new strategy is required.

A number of procedures have already been devised to implement a consistent regularization scheme for inhomogeneous situations, ranging from the simple introduction of an energy cutoff, to relying on the pseudo-potential method to regularize the anomalous density in real space \cite{Bruun-1999}, and to a combination of an energy cutoff with a local-density approximation \cite{BY-2002} (which has then be subject to improvements \cite{LHD-2007}).

In this Appendix, we introduce a procedure to regularize the gap equation (\ref{self-consistency}) under generic inhomogeneous situations, which combines the introduction of an energy cutoff $E_{c}$ as done in subsection II-A for the explicit numerical solution of the BdG equations for eigenvalues $\varepsilon_{\nu}$ up to the value $(E_{c} - \mu$), with the derivation of the Gross-Pitaevskii equation for composite bosons that was done in 
Ref.\cite{PS-2003} in the BEC limit starting from the BdG equations in terms of the small quantity $\Delta(\mathbf{r})/|\mu|$.
Our regularization procedure for the gap equation (\ref{self-consistency}), however, is not limited to the BEC limit but holds instead for any coupling throughout the BCS-BEC crossover, since in the present context it is the ratio $\Delta(\mathbf{r})/(E_{c} - \mu)$ to play the role of the small quantity that allows for the identification of the terms to be retained in the final expression.

Schematically, our regularization procedure of the gap equation (\ref{self-consistency}) is based on the following steps:

\noindent
(i) We consider a wave-vector cutoff $k_{c}$ such that $E_{c}=k_{c}^{2}/(2m)$ is the energy cutoff introduced in subsection II-A.
We take $E_{c} \gg E_{F}$ where $E_{F}=k_{F}^{2}/(2m)$ is the Fermi energy associated with the mean density $n_{0}=k_{F}^{3}/(3 \pi^{2})$.
While $k_{c}$ will be kept finite in the calculation, the ultraviolet cutoff $k_{0} \gg k_{c}$ entering Eq.(\ref{homogeneous-regularization}) will eventually be taken to diverge;

\noindent
(ii) We split the $\sum_{\nu}$ in Eq.(\ref{self-consistency}) in two parts, with $\varepsilon_{\nu} < E_{c} - \mu$ and $\varepsilon_{\nu} > E_{c} - \mu$, respectively.
While in the first part the wave functions $(u_{\nu}(\mathrm{r}),v_{\nu}(\mathrm{r}))$ are explicitly calculated numerically, the second part is treated within a local-density approximation as specified below. In addition, the Fermi function $f_{F}(\epsilon_{\nu})$ can be dropped from this second part for all practical purposes, because $E_{c}/(k_{B}T) \gg 1$ even when $k_{B} T$ is of the order of $E_{F}$;

\noindent
(iii) Following Ref.\cite{BY-2002}, we rewrite the integral on the right-hand side of Eq.(\ref{homogeneous-regularization}) as follows:
\begin{equation}
\int^{k_{0}} \!\!\! \frac{d \mathbf{k}}{(2 \pi)^{3}} \, \frac{m}{\mathbf{k}^{2}} \, \equiv \,
\mathcal{R}(k_{c}) \, + \, \int^{k_{0}}_{k_{c}} \!\!\! \frac{d \mathbf{k}}{(2 \pi)^{3}} \frac{1}{\mathbf{k}^{2}/m - 2 \mu}
\label{definition-R(kc)}
\end{equation}
\noindent
that \emph{defines} the quantity $\mathcal{R}(k_{c})$.
A simple calculation then yields:
\begin{equation}
\begin{small}
\frac{2 \pi^{2}}{m} \, \mathcal{R}(k_{c}) = \left\{
\begin{array}{ll}
k_{c} + \frac{\sqrt{2 m \mu}}{2} \ln \left( \frac{k_{c} - \sqrt{2 m \mu}} {k_{c} + \sqrt{2 m \mu}} \right) & \mbox{($\mu >0$)} \\
k_{c} + \sqrt{2 m |\mu|} \left[ \frac{\pi}{2} - \arctan \left( \frac{k_{c}}{\sqrt{2 m |\mu|}}  \right) \right]   &  \mbox{($\mu <0$)} \, .
\end{array}
\right.
\label{value-R(kc)}
\end{small}
\end{equation}

Through the above steps, the self-consistent condition (\ref{self-consistency}) for the gap parameter becomes:
\begin{eqnarray}
& & \left[ - \frac{m}{4 \pi a_{F}} + \mathcal{R}(k_{c}) + \int^{k_{0}}_{k_{c}} \! \frac{d \mathbf{k}}{(2 \pi)^{3}} \frac{1}{\mathbf{k}^{2}/m - 2 \mu} \right]
\, \Delta(\mathbf{r})  \nonumber \\
& = &  \sum_{\nu}^{\varepsilon_{\nu} < E_{c} - \mu} u_{\nu}(\mathbf{r}) v_{\nu}(\mathbf{r})^{*} \left[ 1 - 2 f_{F}(\epsilon_{\nu}) \right]  \nonumber \\
& + & \sum_{\nu}^{\varepsilon_{\nu} > E_{c} - \mu} u_{\nu}(\mathbf{r}) v_{\nu}(\mathbf{r})^{*} \,\, .
\label{modified-gap-equation-I}
\end{eqnarray}
\noindent
Here, like in Ref.\cite{BY-2002}, the last term within brackets on the left-hand side of Eq.(\ref{modified-gap-equation-I}) can be used to regularize the 
$\sum_{\nu}$ with $\varepsilon_{\nu} > E_{c} - \mu$ on the right-hand side.
What is novel of the present approach, however, is the way this high-energy sum is dealt with, by drawing connections with the derivation of the Gross-Pitaevskii equation for composite bosons in the BEC limit that was done in Ref.\cite{PS-2003} starting from the full BdG equations.

To this end, we refer directly to Eq.(13) of Ref.\cite{PS-2003} and write (by also keeping the same notation of Ref.\cite{PS-2003}):
\begin{eqnarray}
& & \sum_{\nu}^{\varepsilon_{\nu} > E_{c} - \mu} u_{\nu}(\mathbf{r}) v_{\nu}(\mathbf{r})^{*} \cong
\int \! d\mathbf{r_{1}} \, Q(\mathbf{r},\mathbf{r_{1}})^{*} \, \Delta(\mathbf{r_{1}})        \label{equation-13-PS-2003} \\
& + & \int \! d\mathbf{r_{1}} d\mathbf{r_{2}} d\mathbf{r_{3}} \, R(\mathbf{r},\mathbf{r_{1}},\mathbf{r_{2}},\mathbf{r_{3}})^{*} \, \Delta(\mathbf{r_{1}}) \, 
\Delta(\mathbf{r_{2}})^{*} \, \Delta(\mathbf{r_{3}}) \nonumber
\end{eqnarray}
\noindent
where $Q(\mathbf{r},\mathbf{r_{1}})$ and $ R(\mathbf{r},\mathbf{r_{1}},\mathbf{r_{2}},\mathbf{r_{3}})$ are defined by Eqs.(14) and (15) of Ref.\cite{PS-2003}, in the order, but are here considered \emph{with the provision that all} $\mathbf{k}$-\emph{integrals that enter those expressions through the Fourier representation of the non-interacting Green's function} $\tilde{\mathcal{G}}_{0}$ \emph{therein are restricted by} $|\mathbf{k}| > k_{c}$.
In addition, the local-density condition $\mu \rightarrow \mu(\mathbf{r}) = \mu - V_{\mathrm{ext}}(\mathbf{r})$ is adopted here like in Ref.\cite{PS-2003}, to take into account the possible presence of a trapping potential.

Following further Eq.(16) of Ref.\cite{PS-2003}, we approximate for a sufficiently slowly varying gap parameter $\Delta(\mathbf{r})$:
\begin{equation}
\int \! d\mathbf{r_{1}} \, Q(\mathbf{r},\mathbf{r_{1}})^{*} \, \Delta(\mathbf{r_{1}}) \cong \left[ a_{0}(\mathbf{r})^{*} \, + \, 
\frac{1}{2} \, b_{0}(\mathbf{r})^{*} \, \nabla^{2} \right] \, \Delta(\mathbf{r})   \label{equation-16-PS-2003}
\end{equation}
\noindent
where now
\begin{eqnarray}
a_{0}(\mathbf{r}) & \cong & \int_{|\mathbf{k}| > k_{c}} \! \frac{d \mathbf{k}}{(2 \pi)^{3}} \, 
\frac{1}{\mathbf{k}^{2}/m - 2 \mu + 2 V_{\mathrm{ext}}(\mathbf{r})} 
\nonumber \\
& \cong & \int_{|\mathbf{k}| > k_{c}} \! \frac{d \mathbf{k}}{(2 \pi)^{3}} \, \frac{1}{\mathbf{k}^{2}/m - 2 \mu} 
\nonumber \\
& - & 2 V_{\mathrm{ext}}(\mathbf{r}) \int_{|\mathbf{k}| > k_{c}} \! \frac{d \mathbf{k}}{(2 \pi)^{3}} \, 
\frac{1}{\left(\mathbf{k}^{2}/m - 2 \mu\right)^{2}}
\label{approximate-a_0}
\end{eqnarray}
\noindent
and
\begin{eqnarray}
b_{0}(\mathbf{r}) & \cong & \int_{|\mathbf{k}| > k_{c}} \! \frac{d \mathbf{k}}{(2 \pi)^{3}} \, 
\left[ \frac{1}{4m} \frac{1}{ \left( \frac{\mathbf{k}^{2}}{2m} - \mu \right)^{2} } \right.      \nonumber \\
& - & \left. \frac{1}{6m} \frac{ \frac{\mathbf{k}^{2}}{2m}}{ \left( \frac{\mathbf{k}^{2}}{2m} - \mu \right)^{3} } \right] \,\, .
\label{approximate-b_0}
\end{eqnarray}
\noindent
Note that, once these expressions are used in Eq.(\ref{modified-gap-equation-I}), the first term on the right-hand side of Eq.(\ref{approximate-a_0}) cancels the last term within brackets on the left-hand side of Eq.(\ref{modified-gap-equation-I}).
By a similar token we obtain:
\begin{eqnarray}
& & \int \! d\mathbf{r_{1}} d\mathbf{r_{2}} d\mathbf{r_{3}} \, R(\mathbf{r},\mathbf{r_{1}},\mathbf{r_{2}},\mathbf{r_{3}})  \nonumber \\
& \cong &
- \,  \frac{1}{4}  \, \int_{|\mathbf{k}| > k_{c}} \! \frac{d \mathbf{k}}{(2 \pi)^{3}} \, \frac{1}{ \left( \frac{\mathbf{k}^{2}}{2m} - \mu \right)^{3} } \,\, .
 \label{approximate-integral-R}
\end{eqnarray}

Introducing at this point the notation
\begin{equation}
\mathcal{I}_{ij}(k_{c}) \, \equiv \, \int_{|\mathbf{k}|>k_{c}} \! \frac{d\mathbf{k}}{(2 \pi)^{3}} \, 
\frac{\left( \frac{\mathbf{k}^{2}}{2m} \right)^{i}}{\left( \frac{\mathbf{k}^{2}}{2m} - \mu \right)^{j}} \, ,
\label{definition-I_i_j-integrals}
\end{equation}
\noindent
the expression (\ref{equation-13-PS-2003}) can be written compactly as follows:
\begin{eqnarray}
& & \sum_{\nu}^{\varepsilon_{\nu} > E_{c} - \mu} u_{\nu}(\mathbf{r}) v_{\nu}(\mathbf{r})^{*}   \nonumber \\
& \cong & \left[ \int_{|\mathbf{k}|>k_{c}} \! \frac{d \mathbf{k}}{(2 \pi)^{3}} \frac{1}{\mathbf{k}^{2}/m - 2\mu} - 2 V_{\mathrm{ext}}(\mathbf{r})
\, \frac{1}{4} \, \mathcal{I}_{02}(k_{c}) \right] \Delta(\mathbf{r}) 
\nonumber \\
& - & \frac{1}{4} \, \mathcal{I}_{03}(k_{c}) |\Delta(\mathbf{r})|^{2} \Delta(\mathbf{r}) \nonumber \\
& + & \left[ \frac{1}{2} \, \mathcal{I}_{02}(k_{c}) \, - \, \frac{1}{3} \, \mathcal{I}_{13}(k_{c})\right] 
\frac{\nabla^{2} \Delta(\mathbf{r})}{4m} \,\, .
\label{approximate-high_energy-sum}
\end{eqnarray}
\noindent
Here, the first term on the right-hand side which is linear in $\Delta(\mathbf{r})$ was already introduced in Ref.\cite{BY-2002}, while the addition of
the second term on the right-hand side which is cubic in $\Delta(\mathbf{r})$ was already considered in Ref.\cite{LHD-2007}.
What comes naturally from the present derivation is the further introduction of the third term on the right-hand side which emphasizes the spatial variations
of $\Delta(\mathbf{r})$.

Entering the approximate expression (\ref{approximate-high_energy-sum}) into the right-hand side of Eq.(\ref{modified-gap-equation-I}) yields eventually the \emph{regularized gap equation} we were looking for:
\begin{eqnarray}
& & \left\{ - \frac{m}{4 \pi a_{F}} \, + \, \mathcal{R}(k_{c}) \, - \, 
\left[ \frac{1}{2} \, \mathcal{I}_{02}(k_{c}) \, - \, \frac{1}{3} \, \mathcal{I}_{13}(k_{c}) \right] 
\frac{\nabla^{2}}{4m} \right.
\nonumber \\
& &  + \, \left. 2 V_{\mathrm{ext}}(\mathbf{r}) \, \frac{1}{4} \, \mathcal{I}_{02}(k_{c}) \,
+ \, \frac{1}{4} \, \mathcal{I}_{03}(k_{c}) |\Delta(\mathbf{r})|^{2} \right\} \Delta(\mathbf{r})
\nonumber \\
& & = \sum_{\nu}^{\varepsilon_{\nu} < E_{c} - \mu} u_{\nu}(\mathbf{r}) v_{\nu}(\mathbf{r})^{*}  
\left[ 1 - 2 f_{F}(\varepsilon_{\nu}) \right] \,\, .
\label{modified-gap-equation-II}
\end{eqnarray}
\noindent
Note that the expression on the right-hand side, which results from an explicit numerical integration of the BdG equations, acts as a \emph{source term}
on the non-linear differential equation for $\Delta(\mathbf{r})$ given by the left-hand side.

But for the source term on its right-hand side, Eq.(\ref{modified-gap-equation-II}) resembles the Gross-Pitaevskii equation with suitable coefficients, and actually reduces to it in the BEC limit when $\mu \, (<0)$ is the largest energy scale in the problem, such that $E_{c}$ (and thus $k_{c}$) can be taken to vanish for all practical purposes.
In particular, in this limit one obtains for the integrals entering Eq.(\ref{modified-gap-equation-II}) the values:
\begin{eqnarray}
& & \mathcal{R}(k_{c}) \rightarrow \frac{m}{4 \pi a_{F}} \, - \, \frac{m^{2} a_{F}}{8 \pi} \, \mu_{B} 
\,\,\,\, , \,\,\,\, \mathcal{I}_{02}(k_{c}) \rightarrow \frac{m^{2} a_{F}}{2 \pi } \,\,\,\, , \,\,\,\,
\nonumber \\
& & \mathcal{I}_{13}(k_{c}) \rightarrow \frac{3 m^{2} a_{F}}{8 \pi } \,\,\,\, , \,\,\,\, \mathcal{I}_{03}(k_{c}) \rightarrow \frac{m^{3} a_{F}^{3}}{4 \pi } \,\,\,\, ,
\label{integrals-BEC_limit}
\end{eqnarray}
\noindent
where $\mu_{B} = 2 \mu + (m a_{F}^{2})^{-1}$ is the chemical potential for composite bosons.
With the rescaling given in Ref.\cite{PS-2003}, between the gap function $\Delta(\mathbf{r})$ and the condensate wave function for composite bosons,
one recovers in this way from Eq.(\ref{modified-gap-equation-II}) the Gross-Pitaevskii equation given by Eq.(20) of Ref.\cite{PS-2003}.

\begin{figure}[t]
\includegraphics[angle=0,width=8.0cm]{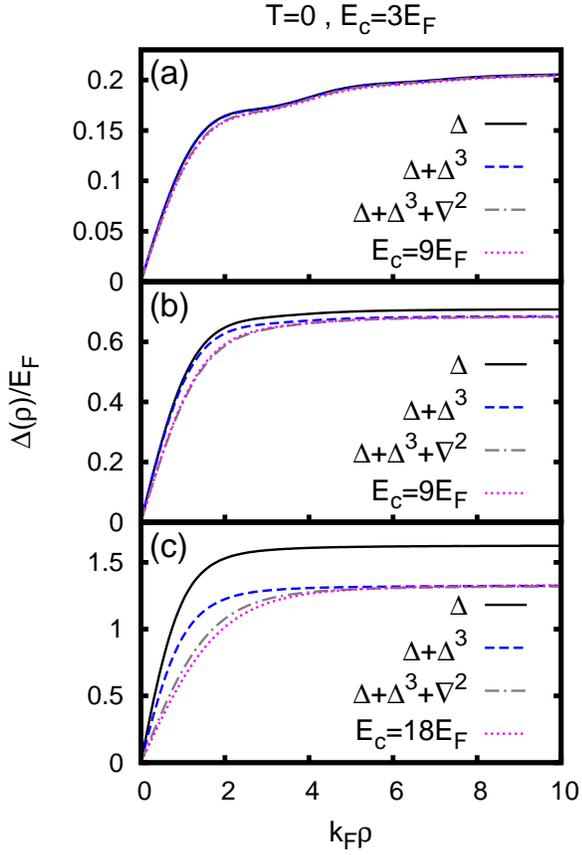}
\caption{$\Delta(\rho)$ (in units of $E_{F}$) vs $\rho$ (in units of $k_{F}^{-1}$) obtained by various approximations at zero temperature for: (a) $(k_{F} a_{F})^{-1}=-1$; (b) $(k_{F} a_{F})^{-1}=0$; (c) $(k_{F} a_{F})^{-1}=+1$.
Comparison is made between the ``best'' calculation with a large value of $E_{c}$ [that equals $9 E_{F}$ in panels (a)
and (b), and $18 E_{F}$ in panel (c)], and less sophisticated calculations all with the smaller value $E_{c} = 3 E_{F}$ which include, respectively, the linear, the linear plus cubic, and the linear plus cubic plus Laplacian terms on the right-hand side of Eq.(\ref{approximate-high_energy-sum}).}
\label{Figure-14}
\end{figure}

In practice, the inclusion of successively more terms on the right-hand side of Eq.(\ref{approximate-high_energy-sum}) [from the linear ($\Delta$) term, to the linear plus cubic ($\Delta + \Delta^{3}$) terms, and finally to the linear plus cubic plus Laplacian ($\Delta + \Delta^{3} + \nabla^{2}$) terms] enables one to decrease the total computational time at any coupling by decreasing the value of the cutoff $E_{c}$ up to which the eigenfunctions of the BdG equations have to be explicitly calculated.
This can be achieved without loosing accuracy in the shape of $\Delta(\mathbf{r})$ as well as of other physical quantities (see also Appendix C).

As an example, we consider again the problem of an isolated vortex in an otherwise infinite superfluid, which is the main concern of the present paper.
For this case, the profile of $\Delta(\rho)$ vs $\rho$ at zero temperature for three different couplings across the BCS-BEC crossover is shown in Fig.÷\ref{Figure-14}, where alternative numerical approximations (including the linear, linear plus cubic, and linear plus cubic plus Laplacian terms), which all adopt a common and rather small value of the cutoff $E_{c}$, are compared with the full calculation of the BdG equations where the value of the cutoff $E_{c}$ is taken considerably larger.
In this case, the smaller value $E_{c} = 3 E_{F}$ of the cutoff with respect to the value $E_{c} = 9 E_{F}$ (or $E_{c} = 18 E_{F}$, depending on the coupling) needed to the full calculation for achieving a stable configuration, yields a reduction of the total computational time by a factor of five or more.
 
Specifically, one sees from Fig.÷\ref{Figure-14} that inclusion of all terms on the right-hand side of Eq.(\ref{approximate-high_energy-sum}) [namely, the linear plus cubic plus Laplacian ($\Delta + \Delta^{3} + \nabla^{2}$) terms] leads for all couplings to quite a good agreement with the full calculation, and not only in the asymptotic (bulk) region but also near the center of the vortex where $\Delta(\rho)$ is strongly depressed.
In contrast, the approximation that includes only the first two terms on the right-hand side of Eq.(\ref{approximate-high_energy-sum}) [namely, the linear plus cubic ($\Delta + \Delta^{3}$) terms] reproduces the bulk value $\Delta_{0}$ but leads to (even sizable) deviations from the full calculation near the center of the vortex.
Finally, the approximation that includes only the first term on the right-hand side of Eq.(\ref{approximate-high_energy-sum}) [namely, the linear ($\Delta$) term] progressively deviates for all values of $\rho$ from the full calculation when approaching the BEC limit, where it is not able to recover the correct bulk value $\Delta_{0}$ with the required accuracy \cite{footnote-Appendix-B}.

\appendix
\section{APPENDIX C: ASYMPTOTIC FORM OF THE NUMBER AND CURRENT DENSITIES}
\label{sec:appendix_C}

Besides the gap $\Delta(\mathbf{r})$, other relevant physical quantities obtained by solving the BdG equations (\ref{BdG-equations}) are the number 
$n(\mathbf{r})$ and current $\mathbf{j}(\mathbf{r})$ densities. 
They are given, respectively, by the expressions:
\begin{small}
\begin{eqnarray}
n(\mathbf{r}) & = & 2 \sum_{\nu} \left[ f_{F}(\epsilon_{\nu}) |u_{\nu}(\mathbf{r})|^{2}  
+ \left( 1 - f_{F}(\epsilon_{\nu}) \right) |v_{\nu}(\mathbf{r})|^{2} \right]                        \label{BdG-density} \\
\mathbf{j}(\mathbf{r}) & = & \frac{1}{i m} \, \sum_{\nu} \left\{
f_{F}(\epsilon_{\nu}) \left[ u_{\nu}(\mathbf{r})^{*} \nabla u_{\nu}(\mathbf{r}) -
(\nabla u_{\nu}(\mathbf{r})^{*}) u_{\nu}(\mathbf{r}) \right]   \right.                              \nonumber \\
& + & \left. \left( 1 - f_{F}(\epsilon_{\nu}) \right)
\left[ v_{\nu}(\mathbf{r}) \nabla v_{\nu}(\mathbf{r})^{*} -
(\nabla v_{\nu}(\mathbf{r})) v_{\nu}(\mathbf{r})^{*} \right]  \right\} .                             \label{BdG-current}
\end{eqnarray}
\end{small}
\noindent
Only positive eigenvalues can be considered for the sums in Eqs.(\ref{BdG-density}) and (\ref{BdG-current}).

In order to calculate the expressions (\ref{BdG-density}) and (\ref{BdG-current}) in an efficient way, and consistently with what was done in Appendix B 
for the gap equation (\ref{self-consistency}), also the $\sum_{\nu}$ in Eqs.(\ref{BdG-density}) and (\ref{BdG-current}) is split into two parts, with $\varepsilon_{\nu} < E_{c} - \mu$ and $\varepsilon_{\nu} > E_{c} - \mu$.
While in the first part with $\varepsilon_{\nu} < E_{c} - \mu$ one uses explicitly the wave functions $(u_{\nu}(\mathrm{r}),v_{\nu}(\mathrm{r}))$ obtained by solving the BdG equations, for both quantities $n(\mathbf{r})$ and $\mathbf{j}(\mathbf{r})$ the second (\emph{asymptotic}) part with $\varepsilon_{\nu} > E_{c} - \mu$ is treated again within a local-density approximation as follows.

\begin{figure}[t]
\includegraphics[angle=0,width=7.5cm]{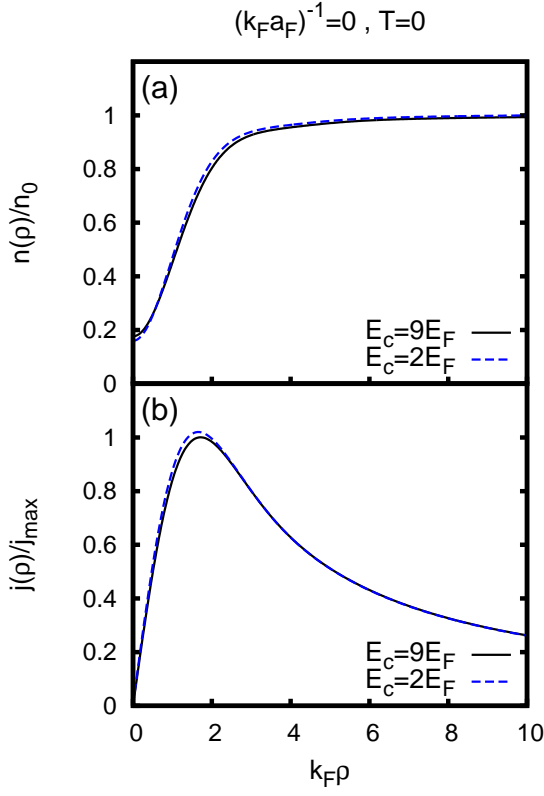}
\caption{(a) Number density $n(\rho)$ (normalized to its bulk value $n_{0}$) and (b) current density $j(\rho)$ (normalized to its maximum value at $R_{\mathrm{v}}$) vs $\rho$ (in units of $k_{F}^{-1}$) obtained at zero temperature and unitarity for an isolated vortex.
Calculations corresponding to two values of the cutoff $E_{c}$ are compared with each other.}
\label{Figure-15}
\end{figure}

Similarly to what was done in Appendix B, we adapt to the present situation the treatment made in Ref.\cite{PS-2003}, where expressions for $n(\mathbf{r})$ and $\mathbf{j}(\mathbf{r})$ consistent with the Gross-Pitaevskii equation were recovered in the BEC limit.
We thus obtain for the \emph{asymptotic} parts:
\begin{equation}
n_{\mathrm{asym}}(\mathbf{r}) \cong \sum_{\nu}^{\varepsilon_{\nu} > E_{c} - \mu} |v_{\nu}(\mathbf{r})|^{2} \cong \frac{1}{2} \, \mathcal{I}_{02}(k_{c}) 
\, |\Delta(\mathbf{r})|^{2}           
\label{asymptotic-density}
\end{equation}
\noindent
and
\begin{eqnarray}
\mathbf{j}_{\mathrm{asym}}(\mathbf{r}) & \cong & \frac{1}{i m} \sum_{\nu}^{\varepsilon_{\nu} > E_{c} - \mu} 
\left[ v_{\nu}(\mathbf{r}) \nabla v_{\nu}(\mathbf{r})^{*} - (\nabla v_{\nu}(\mathbf{r})) v_{\nu}(\mathbf{r})^{*} \right]   
\nonumber \\       
& \cong & \frac{1}{2 i m} \, \left( \frac{1}{2} \, \mathcal{I}_{02}(k_{c}) - \frac{1}{3} \, \mathcal{I}_{03}(k_{c}) \right)                   
\nonumber \\ 
& \times & \left[ \Delta(\mathbf{r})^{*} \nabla \Delta(\mathbf{r}) - \Delta(\mathbf{r}) \nabla \Delta(\mathbf{r})^{*} \right] 
\label{asymptotic-current-density}
\end{eqnarray}
\noindent
with the definition (\ref{definition-I_i_j-integrals}) for the integrals over the wave vector $\mathbf{k}$ with $|\mathbf{k}| > k_{c}$.

The above expressions hold for any coupling. In particular, in the BEC limit, whereby the integrals $\mathcal{I}_{02}(k_{c})$ and $\mathcal{I}_{03}(k_{c})$
reduce to the values (\ref{integrals-BEC_limit}), the expressions (\ref{asymptotic-density}) and (\ref{asymptotic-current-density}) reduce to Eqs.(21) and (22)
of Ref.\cite{PS-2003}, in the order, once the proper rescaling $\Phi(\mathbf{r}) = \sqrt{m^{2} a_{F}/(8 \pi)} \, \Delta(\mathbf{r})$ between the gap function and the condensate wave function $\Phi(\mathbf{r})$ for composite bosons is performed.

\begin{figure}[t]
\includegraphics[angle=0,width=7.5cm]{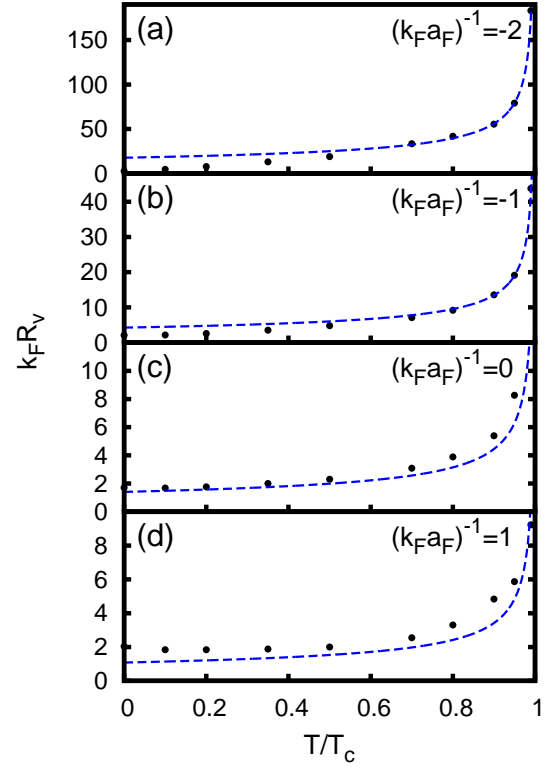}
\caption{The values of the radius $R_{\mathrm{v}}$ of the vortex (dots) are shown versus the temperature $T$ (in units of the respective critical temperature $T_{c}$)
for four different couplings across unitarity.
These values are then fitted by the mean-field-like expression $R_{\mathrm{v}}(T) \propto (T_{c} - T)^{-1/2}$ over the whole temperature range down to $T=0$ 
(dashed lines).}
\label{Figure-16}
\end{figure}

The above expressions can again be applied to the problem of an isolated vortex in an otherwise infinite superfluid.
In particular, in Fig.÷\ref{Figure-15} we show the results of our calculation for the number density $n(\rho)$ and the current density $j(\rho)$ at a distance $\rho$ from the center of the vortex, when different values of the cutoff $E_{c}$ are used.
The calculation is done at zero temperature and unitarity.
Once again we verify that, with the complete regularization procedure for the gap equation introduced in Appendix B and here extended to the density and 
current, rather small values of $E_{c}$ are sufficient in practice to obtain results in quite good agreement with our ``best'' calculation where a large value of 
$E_{c}$ is used. 
In addition, the results of Fig.÷\ref{Figure-15} may serve also implicitly to verify that the approximate expressions (\ref{asymptotic-density}) and (\ref{asymptotic-current-density}), respectively for the asymptotic density and current, were obtained in a physically sound manner.

Note finally that, by this kind of plots, the radius $R_{\mathrm{v}}$ of the vortex can be identified as corresponding to the maximum value of the current.
In turn, the value of $R_{\mathrm{v}}$ fixes a length scale which is relevant to reckon the value of $R_{\mathrm{out}}$, that was introduced in subsection II-A and used in Appendix A to enforce the boundary conditions on the radial wave functions 
(typical values of the ratio $R_{\mathrm{out}}/R_{\mathrm{v}}$ taken in the calculations range from about $50$ at low temperature to about $10$ close to 
$T_{c}$ for the couplings we have explored).

The values of $R_{\mathrm{v}}$ obtained in this way are reported in Fig.÷\ref{Figure-16} for four couplings across unitarity versus the temperature T (in units of the respective critical temperature $T_{c}$).
These values are then fitted by the mean-field-like expression $k_{F} R_{\mathrm{v}} = B \, (1 - T/T_{c})^{-1/2}$, obtaining the values
$B = (17.72,4.26,1.41,1.08)$ for the couplings $(k_{F} a_{F})^{-1} = (-2.0,-1.0,0.0,+1.0)$, in the order.
On the average, these values for the pre-factor $B$ are larger by $\sqrt{2}$ than the values for the pre-factor $A$ entering the corresponding expression 
$k_{F} \xi(T) = A \, (1 - T/T_{c})^{-1/2}$ that were reported in Section IV, thus confirming our conclusion made also in Section IV that a single length scale can be extracted from the BdG equations.



\end{document}